\documentclass[aps]{revtex4}
\usepackage{amsfonts}
\usepackage{amsmath}
\usepackage{amssymb,epsf}

\begin{document}

\title{Magnetic brane solutions of Lovelock gravity with nonlinear
electrodynamics}
\author{Seyed Hossein Hendi$^{1,2}$\footnote{email address: hendi@shirazu.ac.ir}, Behzad Eslam Panah$^{1}$\footnote{
email address: behzad$_{-}$eslampanah@yahoo.com} and Shahram
Panahiyan$^{1}$ \footnote{email address: ziexify@gmail.com}}
\affiliation{$^1$ Physics Department and Biruni Observatory,
College of Sciences, Shiraz
University, Shiraz 71454, Iran\\
$^2$ Research Institute for Astronomy and Astrophysics of Maragha (RIAAM),
Maragha, Iran}

\begin{abstract}
In this paper, we consider logarithmic and exponential forms of
nonlinear electrodynamics as a source and obtain magnetic brane
solutions of the Lovelock gravity. Although these solutions have
no curvature singularity and no horizon, they have a conic
singularity with a deficit angle. We investigate the effects of
nonlinear electrodynamics and the Lovelock gravity on the value of
deficit angle and find that various terms of Lovelock gravity do
not affect deficit angle. Next, we generalize our solutions to
spinning cases with maximum rotating parameters in arbitrary
dimensions and calculate the conserved quantities of the
solutions. Finally, we consider nonlinear electrodynamics as a
correction of the Maxwell theory and investigate the properties of
the solutions.
\end{abstract}

\maketitle

\section{Introduction}

Nonsingular solutions are playing an increasingly important role in physics.
The cosmological singularity at the early universe corresponds to an
infinite energy density state and therefore, it may probably be essential to
consider the quantum gravity to understand the initial state of the
Universe. Hence from the cosmological point of view, nonsingular models of
the Universe have special position for scientists \cite{Nonsingular}. From
gravitational viewpoint, various regular solutions, such as gravitational
instantons, solitons and horizonless magnetic branes (string) solutions,
have become the subject of interest in recent years \cite%
{Instanton,Hirschmann,Dias,MagneticString,MagneticBranes,MagneticStringBI,MagneticBraneBI,MagneticStringPMI,MagneticBranePMI,HendiAHEP}%
.

On the other hand, considering four/higher dimensional spacetimes, the
cosmic strings/branes are topological defects which are inevitably formed
during phase transitions in the early universe \cite{Nielsen}. Investigation
of the horizonless magnetic solutions and their relations to the topological
defects helps us to think about the origin of cosmic magnetic fields \cite%
{Vilenkin,Banerjee}. Besides, from geometric point of view, these structures
are fascinating objects, in which have no curvature singularity, no horizon,
but they have a conic singularity. One of the important motivations for
investigating the horizonless magnetic stings/branes comes from the fact
that these kinds of solutions may be interpreted as cosmic strings/branes.
The horizonless solutions of Einstein and higher derivative gravity theories
in the (absence) presence of the Maxwell and dilaton fields have been
studied in literature \cite{MagneticString,MagneticBranes}. An extension to
include the nonlinear electrodynamics has also been done \cite%
{MagneticStringBI,MagneticBraneBI,MagneticStringPMI,MagneticBranePMI,HendiAHEP}%
.

The purpose of the present paper is constructing a new class of static and
spinning magnetic brane solutions which produces a longitudinal magnetic
field in the background of anti-de Sitter spacetime. These solutions are the
generalization of the solutions of Ref. \cite{HendiAHEP} to higher
dimensions and higher derivative gravity.

%%%%%%%%%%%%%%%%%%%%%%%%%%%%%%%%%%%%%%%%%%%%%%%%%%%%%%%%%%%%%%%%%%%%%

Derivation of various theories by physicists is for the reason of better
description of phenomena in our Universe. It has been confirmed that most of
phenomena in the nature are inherently chaotic and may be described with
nonlinear theories. In electrodynamics domain, although the Maxwell theory
is in agreement with experimental results, it fails regarding some important
issues such as self energy of point-like charges which motivates us to
regard nonlinear electrodynamics (NED). NED theories may be created from
various viewpoint and motivations. For more explanations of some
motivations, we refer the reader to the following brief examples; solving
the problem of point-like charge self energy, compatible with AdS/CFT
correspondence and string theory frames, understanding the nature of
different complex systems, obtaining more information and insight regarding
to quantum gravity, description of pair creation for Hawking radiation and
the behavior of the compact astrophysical objects such as neutron stars and
pulsars \cite{Chen,Fukuma,Aros}. These are some evidences that motivate one
to consider NED theories.

Through last decades different classes of the nonlinear theories were
introduced \cite%
{Born,Hassaine,Soleng,HendiJHEP,HendiAnn1,HendiAnn2,HendiSheykhi,HendiAllahverdizadeh,NEDModels}%
. Among the NED theories, the so-called Born-Infeld (BI) type theories are
quite special, whose Lagrangians may be originated from the string theory.
It has been shown that the low-energy limit of heterotic string theory in
electrodynamics side leads to a quartic correction of the Maxwell field
strength \cite{Kats}. Moreover, one finds that all order loop corrections
may be summed up as a BI type Lagrangian \cite{BIString,Frad,Fradkin}.
Recently, it has been considered two kinds of BI type Lagrangians to examine
the possibility of black hole solutions \cite%
{Soleng,HendiJHEP,HendiAnn1,HendiAnn2,HendiSheykhi,HendiAllahverdizadeh}.
Although there are some analogues between the BI type theories, one can find
that there exist some differences between them.

%%%%%%%%%%%%%%%%%%%%%%%%%%%%%%%%%%%%%%%%%%%%%%%%%%%%%%%%%%%%%%%%%

In recent years a renewed interest has grown in higher dimensional spacetime
as well as higher dimensional gravity \cite{Horava}. The main reason comes
from the fact that these theories emerge in the effective low-energy action
of string theory in gravitational side \cite{Kan,Callen,Horowitz,Gross}. One
of the special classes of higher derivative modifications of Einstein (EN)
gravity is the Lovelock theory \cite{Lovelock} which is a ghost free model
\cite{Callen1,Myers}. Regarding the postulates of general relativity, most
of physicists believe that the Lovelock Lagrangian is a natural
generalization of the EN gravity to higher dimensions. Besides, Lovelock
gravity may solve some of problems of the Einstein theory such as the
normalization problem, and hence it is a well-defined model \cite%
{Stelle,Farhoudi,Witt}. In this paper, we consider the Lovelock gravity in
presence of two classes of BI type NED models and obtain their horizonless
solutions. We also investigate the effect of NED as a correction to the
Maxwell theory.

%%%%%%%%%%%%%%%%%%%%%%%%%%%%%%%%%%%%%%%%%%%%%%%%%%%%%%%%%%%%%%%%%%%%%%%%%%%%%%%%%%%%%%%

The layout of this paper will be this: First we introduce the suitable field
equations regarding to the Lovelock gravity coupled with different magnetic
sources that we are interested in. Next, we obtain static solutions for
metric function. Then, we will consider spinning magnetic string and by
employing counterterm method, we calculate conserved quantities. Last
section will be devoted to closing remarks.

%%%%%%%%%%%%%%%%%%%%%%%%%%%%%%%%%%%%%%%%%%%%%%%%%%%%%%%%%%%%%%%%%%%%%%%%%%%%%%%%%%%%%%%%

\section{Static solutions}

Recently, Dias and Lemos \cite{Dias} have introduced an interesting
spacetime with magnetic brane interpretation which is horizonless. The
mentioned metric in $d$-dimensions may be written as
\begin{equation}
ds^{2}=-\frac{\rho ^{2}}{l^{2}}dt^{2}+\frac{d\rho ^{2}}{f(\rho )}%
+l^{2}f(\rho )d\phi ^{2}+\frac{\rho ^{2}}{l^{2}}dX^{2},  \label{Metric1}
\end{equation}%
where $dX^{2}=\sum_{i=1}^{d_{3}}dx_{i}^{2}$ is the Euclidean metric on the $%
d_{3}$-dimensional submanifold (hereafter we denote $(d-i)$ with $d_{i}$).
The angular coordinate $\phi $ is dimensionless and ranges in $[0,2\pi ]$,
while $x_{i}$ range in $(-\infty ,\infty )$. This metric provides us
horizonless solutions that are of our interest. Now, we are going to obtain
the solutions of first, second and third order of the Lovelock gravity in
the presence of NED with the following field equations
\begin{equation}
\partial _{a}\left( \sqrt{-g}L_{F}F^{ab}\right) =0,  \label{FE1}
\end{equation}%
\begin{equation}
\Lambda g_{ab}+G_{ab}^{(1)}+\alpha _{2}G_{ab}^{(2)}+\alpha _{3}G_{ab}^{(3)}=%
\frac{1}{2}g_{ab}L(F)-2L_{F}F_{ac}F_{b}^{c},  \label{FE2}
\end{equation}%
where $L_{F}=\frac{dL(F)}{dF}$, in which $L(F)$ is the Lagrangian of NED, $%
\Lambda =-\frac{d_{1}d_{2}}{2l^{2}}$ and $G_{ab}^{(1)}=R_{ab}-\frac{1}{2}%
g_{ab}R$ are, respectively, the cosmological constant and the Einstein
tensor, $\alpha _{i}$'s are the Lovelock coefficients and%
\begin{equation}
G_{\mu \nu }^{(2)}=2(R_{\mu \sigma \kappa \tau }R_{\nu }^{\phantom{\nu}%
\sigma \kappa \tau }-2R_{\mu \rho \nu \sigma }R^{\rho \sigma }-2R_{\mu
\sigma }R_{\phantom{\sigma}\nu }^{\sigma }+RR_{\mu \nu })-\frac{\mathcal{L}%
^{(2)}}{2}g_{\mu \nu },  \label{G2}
\end{equation}%
\begin{eqnarray}
G_{\mu \nu }^{(3)} &=&-3(4R^{\tau \rho \sigma \kappa }R_{\sigma \kappa
\lambda \rho }R_{\phantom{\lambda }{\nu \tau \mu}}^{\lambda }-8R_{%
\phantom{\tau \rho}{\lambda \sigma}}^{\tau \rho }R_{\phantom{\sigma
\kappa}{\tau \mu}}^{\sigma \kappa }R_{\phantom{\lambda }{\nu \rho \kappa}%
}^{\lambda }+2R_{\nu }^{\phantom{\nu}{\tau \sigma \kappa}}R_{\sigma \kappa
\lambda \rho }R_{\phantom{\lambda \rho}{\tau \mu}}^{\lambda \rho }  \notag \\
&&-R^{\tau \rho \sigma \kappa }R_{\sigma \kappa \tau \rho }R_{\nu \mu }+8R_{%
\phantom{\tau}{\nu \sigma \rho}}^{\tau }R_{\phantom{\sigma \kappa}{\tau \mu}%
}^{\sigma \kappa }R_{\phantom{\rho}\kappa }^{\rho }+8R_{\phantom
{\sigma}{\nu \tau \kappa}}^{\sigma }R_{\phantom {\tau \rho}{\sigma \mu}%
}^{\tau \rho }R_{\phantom{\kappa}{\rho}}^{\kappa }  \notag \\
&&+4R_{\nu }^{\phantom{\nu}{\tau \sigma \kappa}}R_{\sigma \kappa \mu \rho
}R_{\phantom{\rho}{\tau}}^{\rho }-4R_{\nu }^{\phantom{\nu}{\tau \sigma
\kappa }}R_{\sigma \kappa \tau \rho }R_{\phantom{\rho}{\mu}}^{\rho
}+4R^{\tau \rho \sigma \kappa }R_{\sigma \kappa \tau \mu }R_{\nu \rho
}+2RR_{\nu }^{\phantom{\nu}{\kappa \tau \rho}}R_{\tau \rho \kappa \mu }
\notag \\
&&+8R_{\phantom{\tau}{\nu \mu \rho }}^{\tau }R_{\phantom{\rho}{\sigma}%
}^{\rho }R_{\phantom{\sigma}{\tau}}^{\sigma }-8R_{\phantom{\sigma}{\nu \tau
\rho }}^{\sigma }R_{\phantom{\tau}{\sigma}}^{\tau }R_{\mu }^{\rho }-8R_{%
\phantom{\tau }{\sigma \mu}}^{\tau \rho }R_{\phantom{\sigma}{\tau }}^{\sigma
}R_{\nu \rho }-4RR_{\phantom{\tau}{\nu \mu \rho }}^{\tau }R_{\phantom{\rho}%
\tau }^{\rho }  \notag \\
&&+4R^{\tau \rho }R_{\rho \tau }R_{\nu \mu }-8R_{\phantom{\tau}{\nu}}^{\tau
}R_{\tau \rho }R_{\phantom{\rho}{\mu}}^{\rho }+4RR_{\nu \rho }R_{%
\phantom{\rho}{\mu }}^{\rho }-R^{2}R_{\nu \mu })-\frac{\mathcal{L}^{(3)}}{2}%
g_{\mu \nu },  \label{G3}
\end{eqnarray}%
where $\mathcal{L}^{(2)}$ and $\mathcal{L}^{(3)}$ denote the Lagrangians of
the Gauss-Bonnet (GB) and third order the Lovelock (TOL) gravities, given as
\begin{equation}
\mathcal{L}^{(2)}=R_{\mu \nu \gamma \delta }R^{\mu \nu \gamma \delta
}-4R_{\mu \nu }R^{\mu \nu }+R^{2},  \label{L2}
\end{equation}%
\begin{eqnarray}
\mathcal{L}^{(3)} &=&2R^{\mu \nu \sigma \kappa }R_{\sigma \kappa \rho \tau
}R_{\phantom{\rho \tau }{\mu \nu }}^{\rho \tau }+8R_{\phantom{\mu
\nu}{\sigma \rho}}^{\mu \nu }R_{\phantom {\sigma \kappa} {\nu \tau}}^{\sigma
\kappa }R_{\phantom{\rho \tau}{ \mu \kappa}}^{\rho \tau }+24R^{\mu \nu
\sigma \kappa }R_{\sigma \kappa \nu \rho }R_{\phantom{\rho}{\mu}}^{\rho }
\notag \\
&&+3RR^{\mu \nu \sigma \kappa }R_{\sigma \kappa \mu \nu }+24R^{\mu \nu
\sigma \kappa }R_{\sigma \mu }R_{\kappa \nu }+16R^{\mu \nu }R_{\nu \sigma
}R_{\phantom{\sigma}{\mu}}^{\sigma }-12RR^{\mu \nu }R_{\mu \nu }+R^{3}.
\label{L3}
\end{eqnarray}

In this work, we take into account the recently proposed interesting NED
models \cite{HendiJHEP}. One of them is the Soleng model which is
logarithmic form and another one has exponential form which was proposed by
Hendi with the following explicit forms
\begin{equation}
L(F)=\left\{
\begin{array}{cc}
\beta ^{2}\left[ \exp \left( -\frac{F}{\beta ^{2}}\right) -1\right] & \text{%
ENEF}\vspace{0.1cm} \\
-8\beta ^{2}\ln \left( 1+\frac{F}{8\beta ^{2}}\right) & \text{LNEF}%
\end{array}%
,\right.  \label{L(F)}
\end{equation}%
where $\beta $ is the nonlinearity parameter and the Maxwell invariant is $%
F=F_{ab}F^{ab}$, in which $F_{ab}=\partial _{a}A_{b}-\partial _{b}A_{a}$ is
the electromagnetic field tensor and $A_{a}$ is the gauge potential. It is
easy to show that the electric field comes from the time component of the
vector potential ($A_{t}$), while the magnetic field is associated with the
angular component ($A_{\phi }$). Since we are looking for the magnetic
solutions, we consider the following form of gauge potential%
\begin{equation}
A_{\mu }=h(\rho )\delta _{\mu }^{\phi }.  \label{Amu}
\end{equation}

Using Eq. (\ref{Amu}) with the mentioned NED, one can show that the
electromagnetic field equation (\ref{FE1}) reduces to the following
differential equations
\begin{equation}
\left\{
\begin{array}{cc}
\left( \rho l^{2}\beta ^{2}-4\rho h^{\prime 2}\right) h^{\prime \prime
}+d_{2}l^{2}\beta ^{2}h^{\prime }=0 & \text{ENEF}\vspace{0.1cm} \\
\left( 4\rho l^{2}\beta ^{2}-rh^{\prime 2}\right) h^{\prime \prime
2}+4d_{2}h^{\prime }\left( l^{2}\beta ^{2}+\frac{1}{4}h^{\prime 2}\right) =0
& \text{LNEF}%
\end{array}%
\right. ,  \label{FE11}
\end{equation}%
where the prime and the double prime denote the first and second derivatives
with respect to $\rho $. Solving these equations one obtains
\begin{equation}
h(\rho )=\left\{
\begin{array}{cc}
\frac{l\beta }{2}\int \sqrt{-L_{W1}}d\rho  & \text{ENEF}\vspace{0.1cm} \\
\frac{\beta ^{2}\rho ^{d_{1}}}{qd_{1}}-\frac{\beta ^{2}}{q}\int \Gamma
_{1}\rho ^{d-2}d\rho  & \text{LNEF}%
\end{array}%
\right. ,  \label{h(rho)}
\end{equation}%
where $q$ is an integration constant which is related to the electric
charge, $L_{W1}=LambertW\left( -\left( \frac{4ql}{\beta \rho ^{d_{2}}}%
\right) ^{2}\right) $ and $\Gamma _{1}=\sqrt{1-\left( \frac{2ql}{\beta \rho
^{d_{2}}}\right) ^{2}}$. Taking into account the mentioned gauge potential,
one finds the nonzero components of electromagnetic field are
\begin{equation}
F_{\phi \rho }=-F_{\rho \phi }=\left\{
\begin{array}{cc}
\frac{2ql^{2}}{\rho ^{d_{2}}}\exp \left( -\frac{L_{W1}}{2}\right) , & \text{%
ENEF}\vspace{0.1cm} \\
\frac{\beta ^{2}\rho ^{d_{2}}}{q}\left( 1-\Gamma _{1}\right) , & \text{LNEF}%
\end{array}%
\right. .  \label{FrhoPhi}
\end{equation}

In order to obtain real solutions for the electromagnetic field, we should
restrict the coordinate $\rho $ with a lower bound $\rho _{0}$. It means
\begin{equation*}
\rho >\rho _{0}=\left\{
\begin{array}{cc}
\left( \frac{4ql}{\beta }\right) ^{1/d_{2}}\exp \left( \frac{1}{2d_{2}}%
\right) , & \text{ENEF}\vspace{0.1cm} \\
\left( \frac{2ql}{\beta }\right) ^{1/d_{2}}, & \text{LNEF}%
\end{array}%
\right. .
\end{equation*}

We should note that for large values of $\beta $ all relations reduce to the
corresponding relations of the Maxwell theory. Besides, one can find that
obtained results of electromagnetic fields reduce to those of Ref. \cite%
{HendiAHEP} in four dimensions.

In order to obtain the metric function, $f(\rho )$, one can use nonzero
components of the gravitational field equation, (\ref{FE2}). After
cumbersome calculations, we find that there are two different differential
equations with the following explicit forms
\begin{equation}
e_{t}=\mathcal{K}_{1}+\alpha _{2}\mathcal{K}_{2}+\alpha _{3}\mathcal{K}%
_{3}=0,
\end{equation}%
\begin{equation}
e_{\rho }=\mathcal{K}_{11}+\alpha _{2}\mathcal{K}_{22}+\alpha _{3}\mathcal{K}%
_{33}=0,
\end{equation}%
where%
\begin{equation*}
\mathcal{K}_{1}=-\rho ^{6}\left( \frac{\rho \mathcal{A}^{\prime }}{d_{2}}+%
\mathcal{A}\right) -\beta ^{2}\rho ^{6}\times \left\{
\begin{array}{cc}
1-\exp \left( \frac{-2h^{\prime 2}}{l^{2}\beta ^{2}}\right) , & \text{ENEF}%
\vspace{0.1cm} \\
-8\ln \left( \frac{4l^{2}\beta ^{2}}{4l^{2}\beta ^{2}+h^{\prime 2}}\right) ,
& \text{LNEF}%
\end{array}%
\right. ,
\end{equation*}%
\begin{equation*}
\mathcal{K}_{2}=d_{3}d_{4}\rho ^{4}\left[ 2ff^{\prime \prime }+2f^{\prime 2}+%
\frac{4d_{5}ff^{\prime }}{\rho }+\frac{d_{5}d_{6}f^{2}}{\rho ^{2}}\right] ,
\end{equation*}%
\begin{equation*}
\mathcal{K}_{3}=-d_{3}d_{4}d_{5}d_{6}d_{7}d_{8}f\rho ^{2}\left[ \frac{%
3ff^{\prime \prime }+6f^{\prime 2}}{d_{7}d_{8}}+\frac{6ff^{\prime }}{%
d_{8}\rho }+\frac{f^{2}}{\rho ^{2}}\right] ,
\end{equation*}%
\begin{equation*}
\mathcal{K}_{11}=\rho ^{6}\mathcal{A}-\beta ^{2}\rho ^{6}\times \left\{
\begin{array}{cc}
\frac{4h^{\prime 2}}{l^{2}\beta ^{2}\rho }\exp \left( \frac{-2h^{\prime 2}}{%
l^{2}\beta ^{2}}\right) +\exp \left( \frac{-2h^{\prime 2}}{l^{2}\beta ^{2}}%
\right) -1, & \text{ENEF}\vspace{0.1cm} \\
8\left( \frac{2}{1+\left( \frac{2l\beta }{h^{\prime }}\right) ^{2}}+\ln %
\left[ 1+\left( \frac{h^{\prime }}{2l\beta }\right) ^{2}\right] \right) , &
\text{LNEF}%
\end{array}%
\right. ,
\end{equation*}%
\begin{equation*}
\mathcal{K}_{22}=-d_{2}d_{3}d_{4}d_{5}f\rho ^{2}\left( f+\frac{2\rho
f^{\prime }}{d_{5}}\right) ,
\end{equation*}%
\begin{equation*}
\mathcal{K}_{33}=d_{2}d_{3}d_{4}d_{5}d_{6}d_{7}f^{2}\left( f+\frac{3\rho
f^{\prime }}{d_{7}}\right) .
\end{equation*}%
Now, we desire to obtain higher dimensional magnetic brane solutions in the
EN, GB and TOL gravities, separately. One can set $\alpha _{3}=0$ to obtain
the GB solutions and for $\alpha _{2}=\alpha _{3}=0$, we obtain magnetic
solutions of the EN gravity. After some simplifications, we obtain

%%%%%%%%%%%%%%%%%%%%%%%%%%%%%%%%%%%%%%%%%%%%%%%%%%%%%%%%%%%%%%%
\begin{figure}[tbp]
$%
\begin{array}{cc}
\epsfxsize=7cm \epsffile{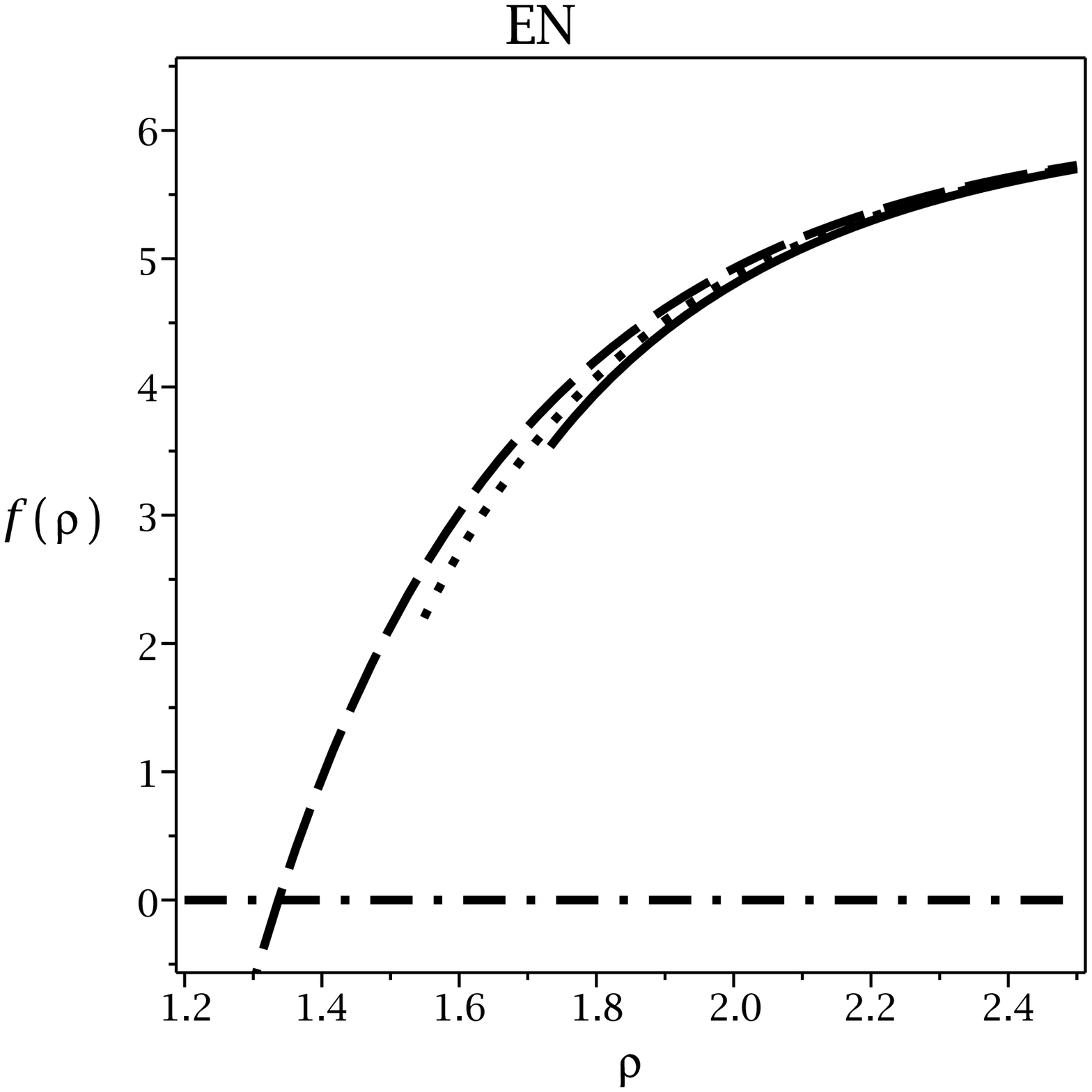} & \epsfxsize=7cm \epsffile{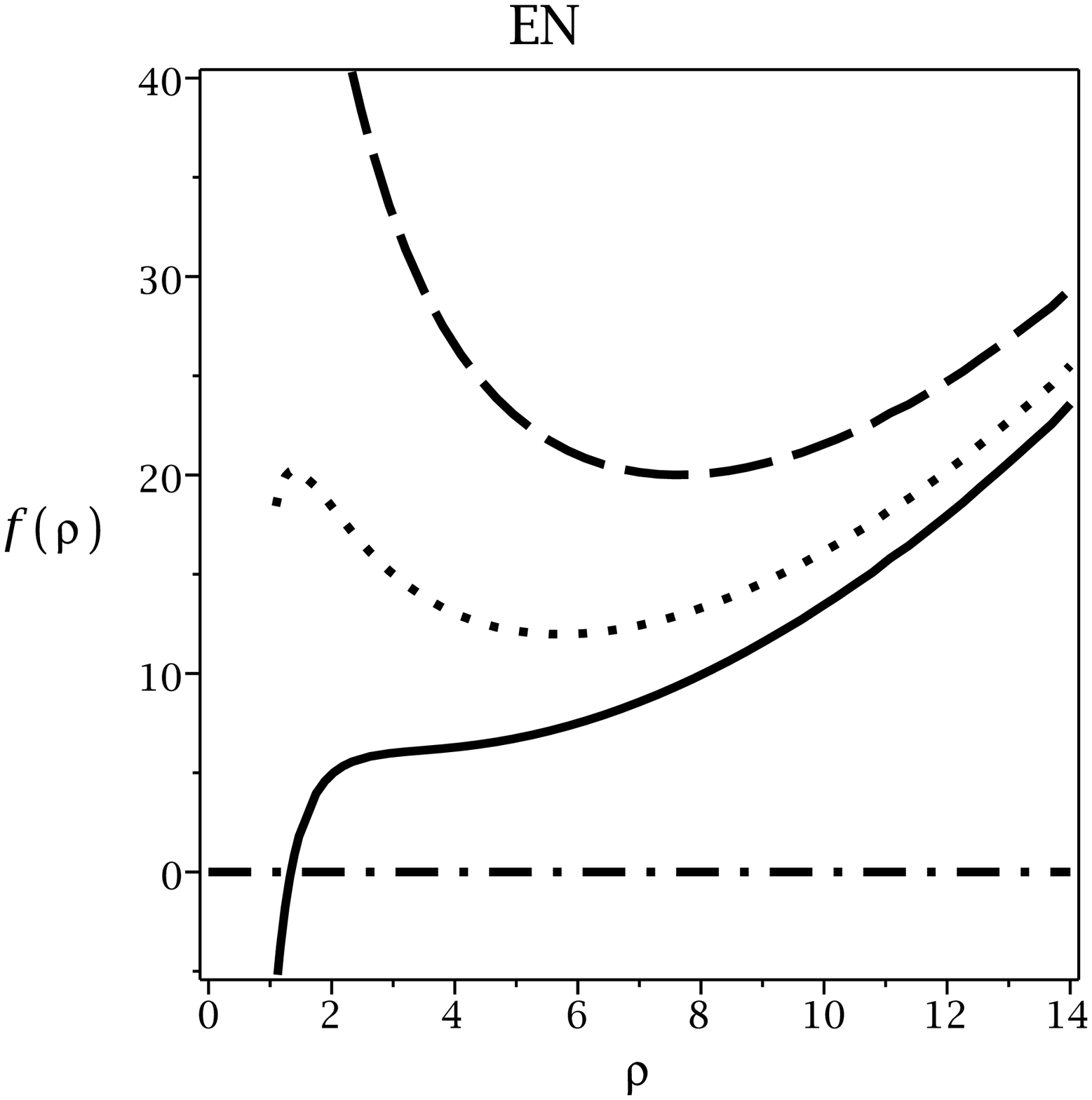}%
\end{array}
$%
\caption{\textbf{LNEF branch of EN gravity:} $f_{EN}(\rho)$ versus $\protect\rho$ for $l=3$%
, $q=1$ and $d=4$. \newline
Left panel: $m=0.5$, $\protect\beta=2$ (continuous line), $\protect\beta=2.5$
(doted line) and $\protect\beta=5$ (dashed line). \newline
Right panel: $\protect\beta=5$, $m=0.5$ (continuous line), $m=1$ (doted
line), $m=2$ (dashed line).}
\label{FigfEN}
\end{figure}

%%%%%%%%%%%%%%%%%%%%%%%%%%%%%%%%%%%%%%%%%%%%%%%%%%%%%%%%%%%%%%%
%%%%%%%%%%%%%%%%%%%%%%%%%%%%%%%%%%%%%%%%%%%%%%%%%%%%%%%%%%%%%%%
\begin{figure}[tbp]
$%
\begin{array}{cc}
\epsfxsize=7cm \epsffile{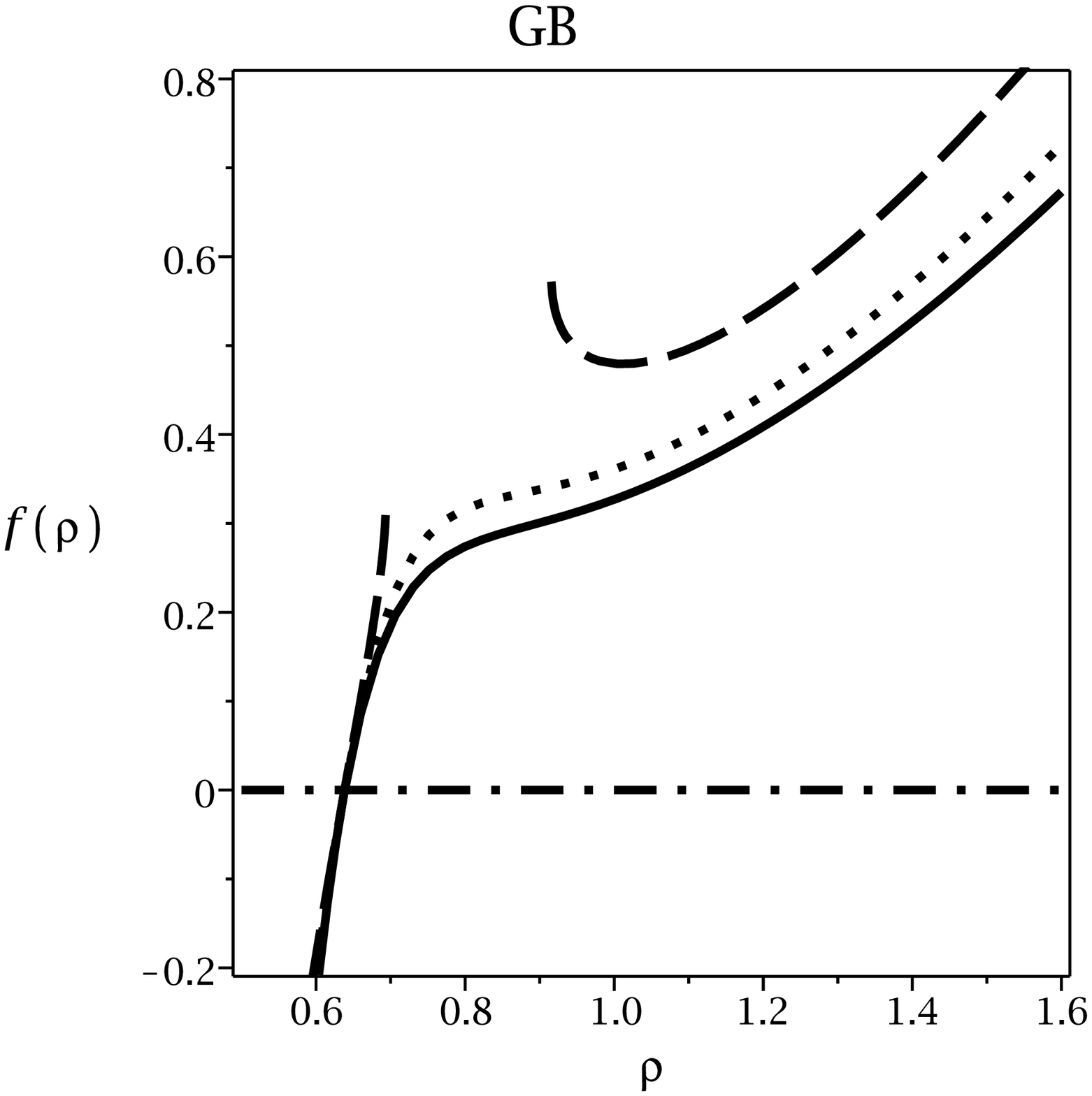} & \epsfxsize=7cm \epsffile{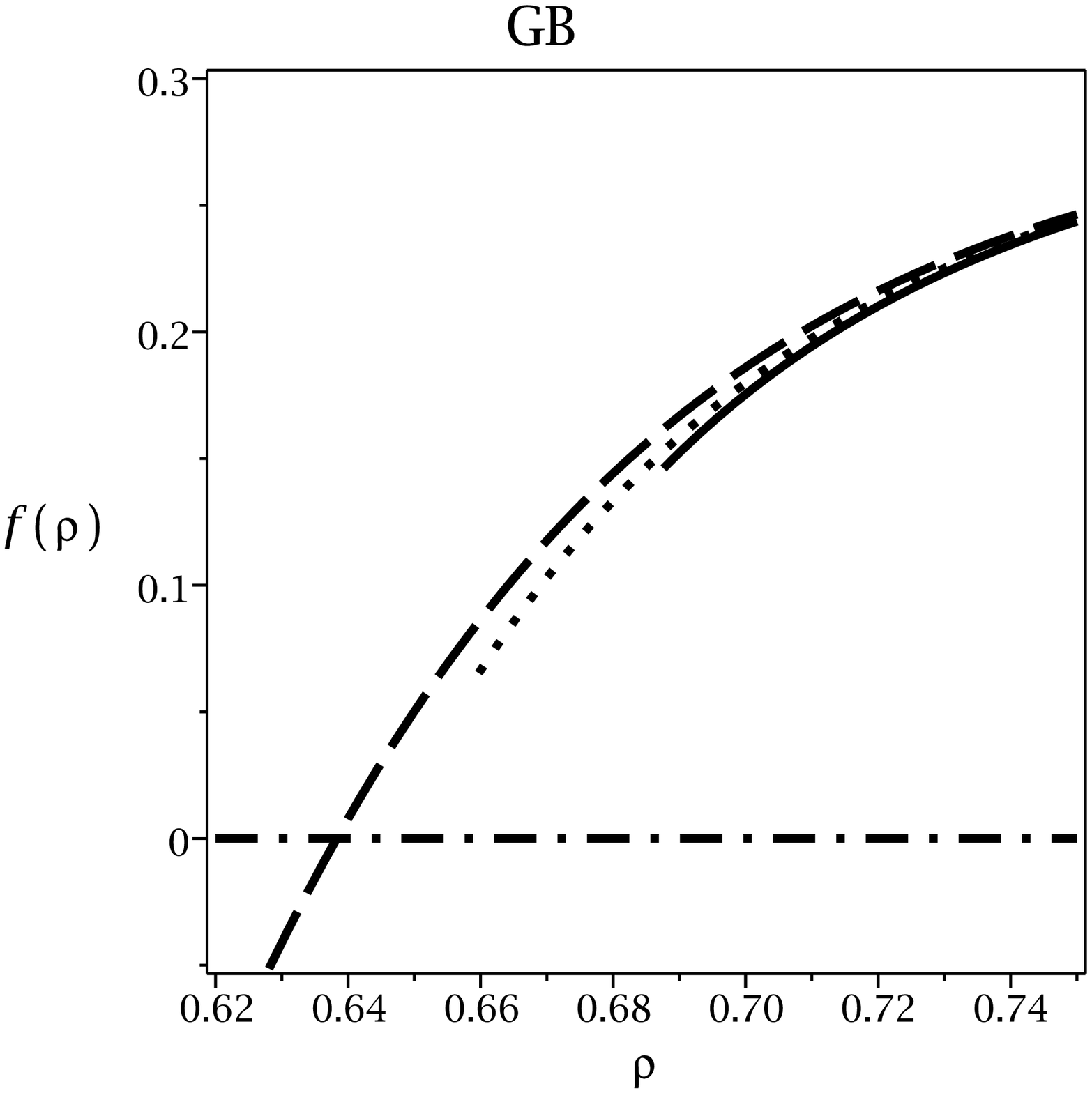}%
\end{array}
$%
\caption{\textbf{LNEF branch of GB gravity:} $f_{GB}(\rho)$ versus $\protect\rho$ for $l=2$%
, $q=0.1$, $m=0.005$ and $d=7$. \newline Left panel:
$\protect\beta=10$, $\protect\alpha_{2}=0.01$ (continuous line),
$\protect\alpha_{2}=0.03$ (doted line) and
$\protect\alpha_{2}=0.06$ (dashed line). \newline Right panel:
$\protect\alpha_{2}=0.01$, $\protect\beta=2.6$ (continuous line),
$\protect\beta=3.2$ (doted line), $\protect\beta=10$ (dashed
line).} \label{FigFGB}
\end{figure}

%%%%%%%%%%%%%%%%%%%%%%%%%%%%%%%%%%%%%%%%%%%%%%%%%%%%%%%%%%%%%%%
%%%%%%%%%%%%%%%%%%%%%%%%%%%%%%%%%%%%%%%%%%%%%%%%%%%%%%%%%%%%%%%
\begin{figure}[tbp]
$%
\begin{array}{cc}
\epsfxsize=7cm \epsffile{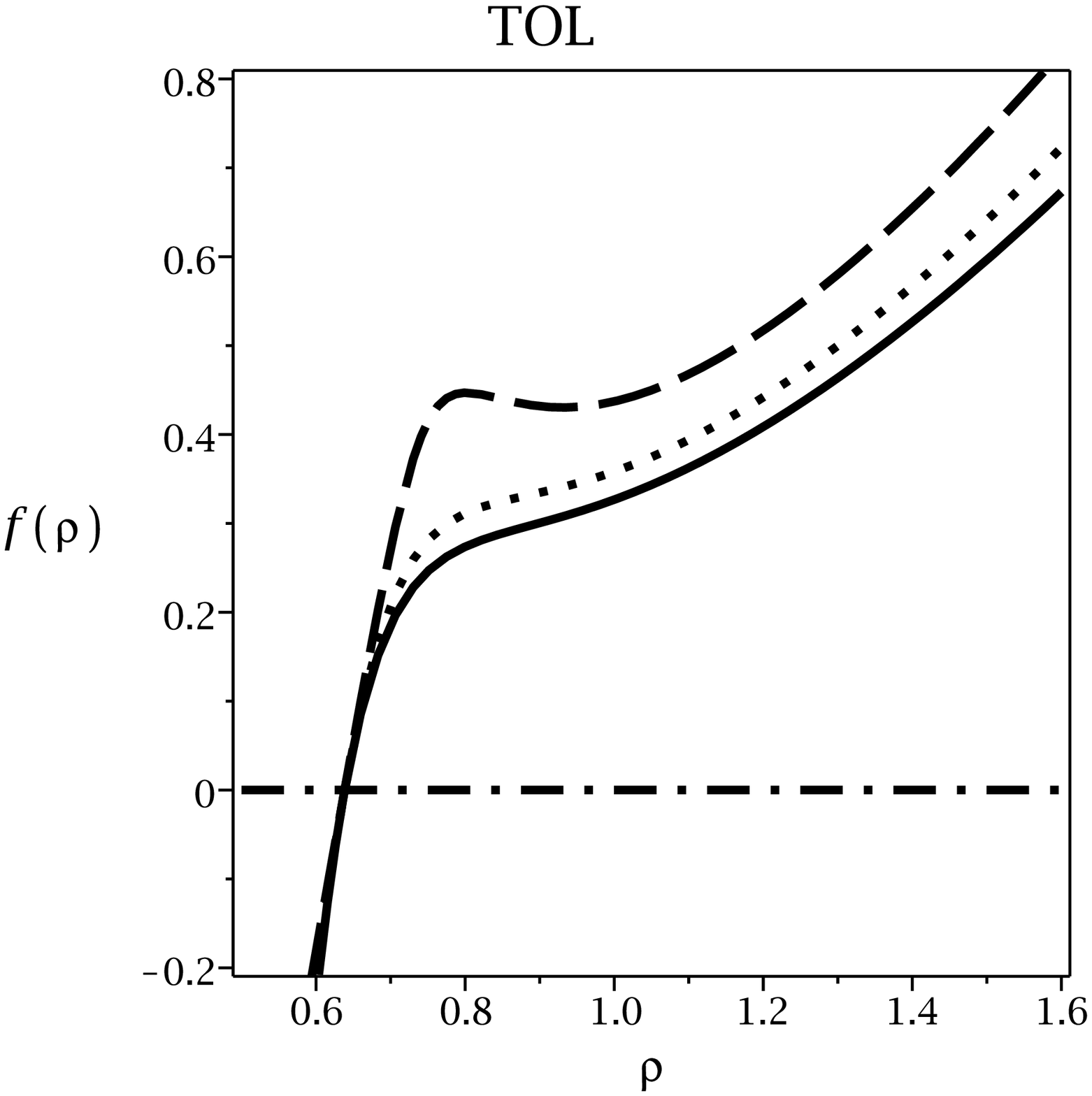} & \epsfxsize=7cm \epsffile{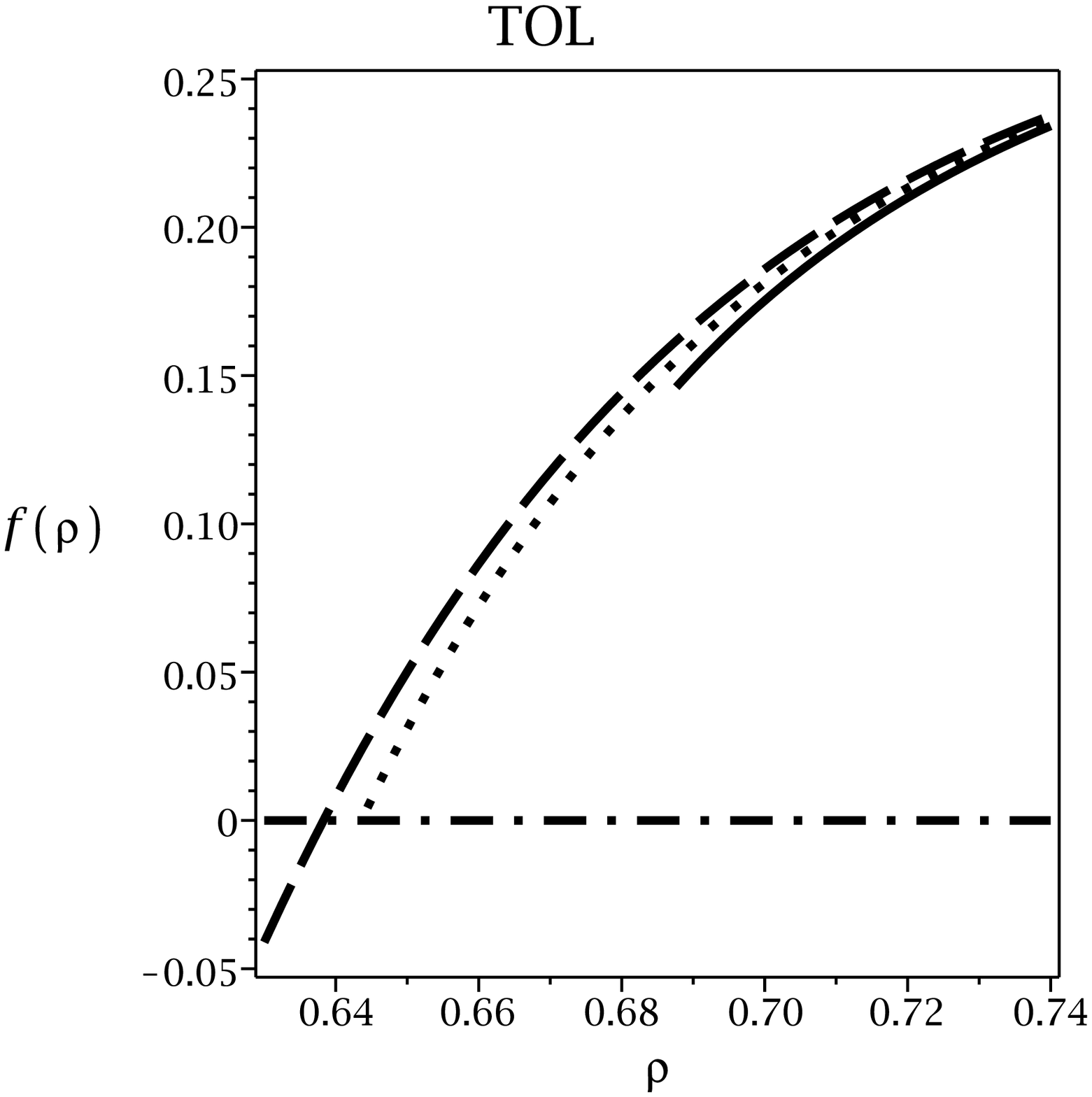}%
\end{array}
$%
\caption{\textbf{LNEF branch of TOL gravity:} $f_{TOL}(\rho)$ versus $\protect\rho$ for $%
l=2 $, $q=0.1$, $m=0.005$ and $d=7$. \newline Left panel:
$\protect\beta=10$, $\protect\alpha_{2}=0.01$ (continuous line),
$\protect\alpha_{2}=0.03$ (doted line) and
$\protect\alpha_{2}=0.06$ (dashed line). \newline Right panel:
$\protect\alpha_{2}=0.01$, $\protect\beta=2.6$ (continuous line),
$\protect\beta=3.2$ (doted line), $\protect\beta=10$ (dashed
line).} \label{FigFTOL}
\end{figure}
%%%%%%%%%%%%%%%%%%%%%%%%%%%%%%%%%%%%%%%%%%%%%%%%%%%%%%%%%%%%%%%

\begin{equation}
f_{EN}=\frac{2ml^{3}}{\rho ^{d_{3}}}-\frac{2\Lambda \rho ^{2}}{d_{1}d_{2}}%
+\left\{
\begin{array}{cc}
\frac{8\beta ^{2}\rho ^{2}}{d_{1}d_{2}}+\frac{-8\beta ^{2}\left( \int \rho
^{d_{2}}\left[ \Gamma _{1}+\ln \left( \frac{\beta ^{2}\rho ^{2d_{2}}\left(
1-\Gamma \right) }{2l^{2}q^{2}}\right) \right] d\rho \right) }{d_{2}\rho
^{d_{3}}} & \text{LNEF} \\
\text{ \ \ }-\frac{\beta ^{2}\rho ^{2}}{d_{1}d_{2}}+\frac{4lq\beta \left(
\int \left[ \sqrt{-L_{W1}}+\frac{1}{\sqrt{-L_{W1}}}\right] d\rho \right) }{%
d_{2}\rho ^{d_{3}}} & \text{ENEF}%
\end{array}%
\right. ,  \label{F1}
\end{equation}%
\begin{equation}
f_{GB}=\frac{\rho ^{2}}{2d_{3}d_{4}\alpha _{2}}\left( 1-\Psi ^{1/2}\right) ,
\label{F2}
\end{equation}%
\begin{equation}
f_{TOL}=\frac{\rho ^{2}}{d_{3}d_{4}\alpha _{2}}\left( 1-\Psi ^{1/3}\right) ,
\label{F3}
\end{equation}%
where%
\begin{equation}
\Psi =1+\frac{2\chi d_{3}d_{4}\alpha _{2}}{d_{1}d_{2}}\left( \Lambda -\frac{%
d_{1}d_{2}l^{3}m}{\rho ^{d_{1}}}+\mathcal{W}\text{\ }\right) ,  \label{PsiGB}
\end{equation}%
with
\begin{equation}
\mathcal{W}=\left\{
\begin{array}{cc}
4\beta ^{2}\left\{ \ln \left( \frac{\beta ^{2}\rho ^{2d_{2}}}{2l^{2}q^{2}}%
\left[ 1-\Gamma _{1}\right] \right) -\frac{(2d_{2}+1)}{d_{1}}\Gamma
_{1}\right\} +\frac{16d_{2}^{2}l^{2}q^{2}}{\rho ^{2d_{2}}d_{1}d_{3}}\mathcal{%
F} & \text{LNEF} \\
\beta ^{2}\left[ \frac{1}{2}+\frac{2d_{1}ql}{\beta \rho ^{d_{1}}}\int \left(
\sqrt{-L_{W1}}+\frac{1}{\sqrt{-L_{W1}}}\right) d\rho \right] \text{ \ \ } &
\text{ENEF}%
\end{array}%
\right. ,
\end{equation}%
in which $\chi =4$ and $3$ for the GB theory and the TOL gravity,
respectively, $\mathcal{F}$ is $_{2}F_{1}\left( \left[ \frac{1}{2},\frac{%
d_{3}}{2d_{2}}\right] ,\left[ \frac{3d_{2}-1}{2d_{2}}\right] ,\frac{%
4l^{2}q^{2}}{\beta ^{2}\rho ^{2d_{2}}}\right) $, $m$\ is an integration
constant related to total finite mass of the solutions and we set $\alpha
_{3}=\frac{d_{3}d_{4}}{3d_{5}d_{6}}\alpha _{2}^{2}$ for more simplifications
of TOL gravity solutions.

\subsection{Properties of solutions:}

At the first step, we are going to discuss the geometric
properties of the solutions. To do this, we look for possible
black hole solutions with obtaining the curvature singularities
and their horizons. We usually calculate the Kretschmann scalar,
$R_{\alpha \beta \gamma \delta }R^{\alpha \beta \gamma \delta }$,
to achieve essential singularity. Considering the mentioned
spacetime, (\ref{Metric1}), it is easy to show that
\begin{equation}
R_{\alpha \beta \gamma \delta }R^{\alpha \beta \gamma \delta }=f^{\prime
\prime 2}+2d_{2}\left( \frac{f^{\prime }}{\rho }\right)
^{2}+2d_{2}d_{3}\left( \frac{f}{\rho ^{2}}\right) ^{2}.  \label{RR}
\end{equation}

Inserting the metric function, $f(\rho )$, in Eq. (\ref{RR}) and using
numerical analysis, one finds that the Kretschmann scalar diverges at $\rho
=\rho_{0}$ and it is finite for $\rho >\rho _{0}$ and naturally one may
think that there is a curvature singularity located at $\rho =\rho_{0}$. In
what follows, we state an important point, in which confirms that the
spacetime never achieves $\rho =\rho_{0}$. As one can confirm, easily, the
metric function has positive value for large values of $\rho >>\rho _{0}$.
So two cases may occur. For the first case, $f(\rho)$ is a positive definite
function with no root and therefore, the singularity called as a naked
singularity which we are not interested in. We consider the second case, in
which the metric function has one or more real positive root(s) larger than $%
\rho _{0}$.

From Figs. \ref{FigfEN} -- \ref{FigFTOL}, we find that there is a $\rho
_{min}$ ($\rho_{min}=\rho_{0}$) in which for $\rho \geq \rho _{min}$ the
metric function is real. These figures show that increasing the nonlinearity
parameter leads to decreasing $\rho _{min}$. Since we are looking for the
metric function with at least one real root, we should adjust the metric
parameters with a suitable range of nonlinearity parameter to obtain $f(\rho
=\rho _{min})\leq 0$.

Moreover, Fig. \ref{FigFTOL} indicates that although metric function of the
TOL gravity is real for arbitrary $\rho $, in GB gravity one encounters with
an imaginary interval for some values of the GB parameter. In other words,
in GB gravity we should adjust the metric parameters with suitable interval
of $\alpha $ to obtain a real metric function with at least one real root.
Besides, Fig. \ref{FigFTOL} shows that the root of metric function does not
depend on the Lovelock parameters.

Now, we denote $r_{+}$ as the largest real positive root of $f(\rho )$. The
metric function is negative for $\rho <r_{+}$ and positive for $\rho >r_{+}$
and hence, the metric signature may change from ($-++++...+$) to ($---++...+$%
) in the range $0<\rho <r_{+}$. Taking into account this apparent change of
signature of the metric, we conclude that one cannot extend the spacetime to
$\rho <r_{+}$. In order to get rid of this incorrect extension, one may use
the following suitable transformation with introducing a new radial
coordinate $r$
\begin{equation}
\begin{array}{c}
r^{2}=\rho ^{2}-r_{+}^{2}, \\
\rho \geq r_{+}\Longleftrightarrow r\geq 0.%
\end{array}
\label{Transformation}
\end{equation}

Using the mentioned transformation with $d\rho =\frac{r}{\sqrt{%
r^{2}+r_{+}^{2}}}dr$ one finds that the metric (\ref{Metric1}) should change
to
\begin{equation}
ds^{2}=-\frac{r^{2}+r_{+}^{2}}{l^{2}}dt^{2}+\frac{r^{2}}{\left(
r^{2}+r_{+}^{2}\right) f(r)}dr^{2}+l^{2}f(r)d\phi ^{2}+\frac{r^{2}+r_{+}^{2}%
}{l^{2}}dX^{2}.  \label{Metric2}
\end{equation}

It is worthwhile to mention that with this new coordinate, the
electromagnetic field and the metric functions lead to the following form%
\begin{equation}
F_{r\phi }=\left\{
\begin{array}{cc}
\frac{2ql^{2}}{\left( r^{2}+r_{+}^{2}\right) ^{\frac{d_{2}}{2}}}\exp \left( -%
\frac{L_{W}}{2}\right) , & \text{ENEF}\vspace{0.1cm} \\
\frac{\beta ^{2}\left( r^{2}+r_{+}^{2}\right) ^{\frac{d_{2}}{2}}}{q}\left(
1-\Gamma \right) , & \text{LNEF}%
\end{array}%
\right. ,  \label{Frphi}
\end{equation}%
\begin{equation}
f_{EN}=\frac{2ml^{3}}{\left( r^{2}+r_{+}^{2}\right) ^{\frac{d_{3}}{2}}}-%
\frac{2\Lambda \left( r^{2}+r_{+}^{2}\right) }{d_{1}d_{2}}+\left\{
\begin{array}{cc}
\frac{8\beta ^{2}\left( r^{2}+r_{+}^{2}\right) }{d_{1}d_{2}}+\frac{-8\beta
^{2}\left( \int r\left( r^{2}+r_{+}^{2}\right) ^{\frac{d_{3}}{2}}\left[
\Gamma +\ln \left( \frac{\beta ^{2}\left( r^{2}+r_{+}^{2}\right) ^{d_{2}}}{%
2l^{2}q^{2}}\left( 1-\Gamma \right) \right) \right] dr\right) }{d_{2}\left(
r^{2}+r_{+}^{2}\right) ^{\frac{d-3}{2}}} & \text{LNEF} \\
\text{ \ \ }-\frac{\beta ^{2}\left( r^{2}+r_{+}^{2}\right) }{d_{1}d_{2}}+%
\frac{4lq\beta \left( \int \left( \sqrt{-L_{W}}+\frac{1}{\sqrt{-L_{W}}}%
\right) \frac{r}{\sqrt{r^{2}+r_{+}^{2}}}dr\right) }{d_{2}\left(
r^{2}+r_{+}^{2}\right) ^{\frac{d_{3}}{2}}} & \text{ENEF}%
\end{array}%
\right. ,  \label{F11}
\end{equation}%
\begin{equation}
f_{GB}=\frac{\left( r^{2}+r_{+}^{2}\right) }{2d_{3}d_{4}\alpha _{2}}\left(
1-\Psi ^{1/2}\right) ,  \label{F22}
\end{equation}%
\begin{equation}
f_{TOL}=\frac{\left( r^{2}+r_{+}^{2}\right) }{d_{3}d_{4}\alpha _{2}}\left(
1-\Psi ^{1/3}\right) ,  \label{F33}
\end{equation}%
where%
\begin{equation}
\Psi =1+\frac{2\chi d_{3}d_{4}\alpha _{2}}{d_{1}d_{2}}\left( \Lambda -\frac{%
d_{1}d_{2}l^{3}m}{\left( r^{2}+r_{+}^{2}\right) ^{d_{1}/2}}+\mathcal{W}%
_{1}\right) ,  \label{PSI23}
\end{equation}%
with
\begin{equation}
\mathcal{W}_{1}=\left\{
\begin{array}{cc}
4\beta ^{2}\left\{ \ln \left( \frac{\beta ^{2}\left( r^{2}+r_{+}^{2}\right)
^{d_{2}}}{2l^{2}q^{2}}\left[ 1-\Gamma \right] \right) -\frac{(2d_{2}+1)}{%
d_{1}}\Gamma \right\} +\frac{16d_{2}^{2}l^{2}q^{2}}{\left(
r^{2}+r_{+}^{2}\right) ^{d_{2}}d_{1}d_{3}}\mathcal{F} & \text{LNEF} \\
\beta ^{2}\left[ \frac{1}{2}+\frac{2d_{1}ql}{\beta \left(
r^{2}+r_{+}^{2}\right) ^{\frac{d_{1}}{2}}}\int \left( \sqrt{-L_{W}}+\frac{1}{%
\sqrt{-L_{W}}}\right) \frac{r}{\sqrt{r^{2}+r_{+}^{2}}}dr\right] \text{ \ \ }
& \text{ENEF}%
\end{array}%
\right. ,
\end{equation}%
in which $L_{W}=LambertW\left( -\frac{16q^{2}l^{2}}{\beta ^{2}\left(
r^{2}+r_{+}^{2}\right) ^{d_{2}}}\right) $, $\mathcal{F}={_{2}F_{1}\left( %
\left[ \frac{1}{2},\frac{d_{3}}{2d_{2}}\right] ,\left[ \frac{3d_{2}-1}{2d_{2}%
}\right] ,\frac{4l^{2}q^{2}}{\beta ^{2}\left( r^{2}+r_{+}^{2}\right) ^{d_{2}}%
}\right) }$ and $\Gamma =\sqrt{1-\frac{4q^{2}l^{2}}{\beta ^{2}\left(
r^{2}+r_{+}^{2}\right) ^{d_{2}}}}$. Since we suppose that $r_{+}\geq \rho
_{0}$, the solutions (electromagnetic field and metric functions) are real
for $r\geq 0$. In addition, the function $f(r)$ given in Eqs. (\ref{F11})-(%
\ref{F33}) is positive in the whole spacetime and is zero at $r=0$.

Although the Kretschmann scalar does not diverge in the range $0\leq
r<\infty $, one can show that there is a conical singularity at $r=0$. One
can investigate the conic geometry by using the \textit{circumference/radius}
ratio. Using the Taylor expansion, in the vicinity of $r=0$, we find
\begin{equation}
f(r)=\left. f(r)\right\vert _{r=0}+\left( \left. \frac{df(r)}{dr}\right\vert
_{r=0}\right) r+\frac{1}{2}\left( \left. \frac{d^{2}f(r)}{dr^{2}}\right\vert
_{r=0}\right) r^{2}+O(r^{3})+...,
\end{equation}%
where
\begin{equation*}
\left. f(r)\right\vert _{r=0}=\left. \frac{df(r)}{dr}\right\vert _{r=0}=0,
\end{equation*}%
and it is a matter of calculation to show that regardless of gravity
branches (EN, GB and TOL), we will have following relation
\begin{equation}
\left. \frac{d^{2}f(r)}{dr^{2}}\right\vert _{r=0}=-\frac{2\Lambda }{d_{2}}+%
\frac{2}{d_{2}}E_{0}+\frac{2r_{+}}{d_{1}d_{2}}E_{0}^{\prime }\neq 0,
\label{f''}
\end{equation}%
where $E_{0}=\left. E(r)\right\vert _{r=0}$, in which $E(r)$ denotes the
electromagnetic part of metric functions (third term of Eq. (\ref{F11}) and $%
\mathcal{W}_{1}$ in Eq. (\ref{PSI23}), and $E_{0}^{\prime }=\left. \frac{%
dE(r)}{dr}\right\vert _{r=0}$. With employing obtained results,
one can show that
\begin{equation}
\lim_{r\longrightarrow 0^{+}}\frac{1}{r}\sqrt{\frac{g_{\phi \phi }}{g_{rr}}}%
=\lim_{r\longrightarrow 0^{+}}\frac{\sqrt{r^{2}+r_{+}^{2}}lf(r)}{r^{2}}=%
\frac{lr_{+}}{2}\left. \frac{d^{2}f(r)}{dr^{2}}\right\vert _{r=0}\neq 1,
\label{lim1}
\end{equation}%
which confirms that as the radius $r$ tends to zero, the limit of the
\textit{circumference/radius} ratio is not $2\pi $ and therefore the
spacetime has a conical singularity at $r=0$. This canonical singularity may
be removed if one identifies the coordinate $\phi $ with the period%
\begin{equation}
\text{Period}_{\phi }=2\pi \left( \lim_{r\longrightarrow 0}\frac{1}{r}\sqrt{%
\frac{g_{\phi \phi }}{g_{rr}}}\right) ^{-1}=2\pi \left( 1-4\mu \right) ,
\end{equation}%
where $\mu $ is given by%
\begin{equation}
\mu =\frac{1}{4}\left[ 1-\frac{2}{lr_{+}}\left( \left. \frac{d^{2}f(r)}{%
dr^{2}}\right\vert _{r=0}\right) ^{-1}\right] .  \label{mu}
\end{equation}

In other words, the near origin limit of the metric (\ref{Metric2})
describes a locally flat spacetime which has a conical singularity at $r=0$
with a deficit angle $\delta \phi =8\pi \mu $. Using the Vilenkin procedure,
one can interpret $\mu $ as the mass per unit volume of the magnetic brane
\cite{Vilenkin1985}. It is evident from (\ref{f''}) and (\ref{mu}) that
deficit angle is independent of the Lovelock coefficients and is only a
function of cosmological constant and electromagnetic field.

\begin{figure}[tbp]
$%
\begin{array}{ccc}
\epsfxsize=7cm \epsffile{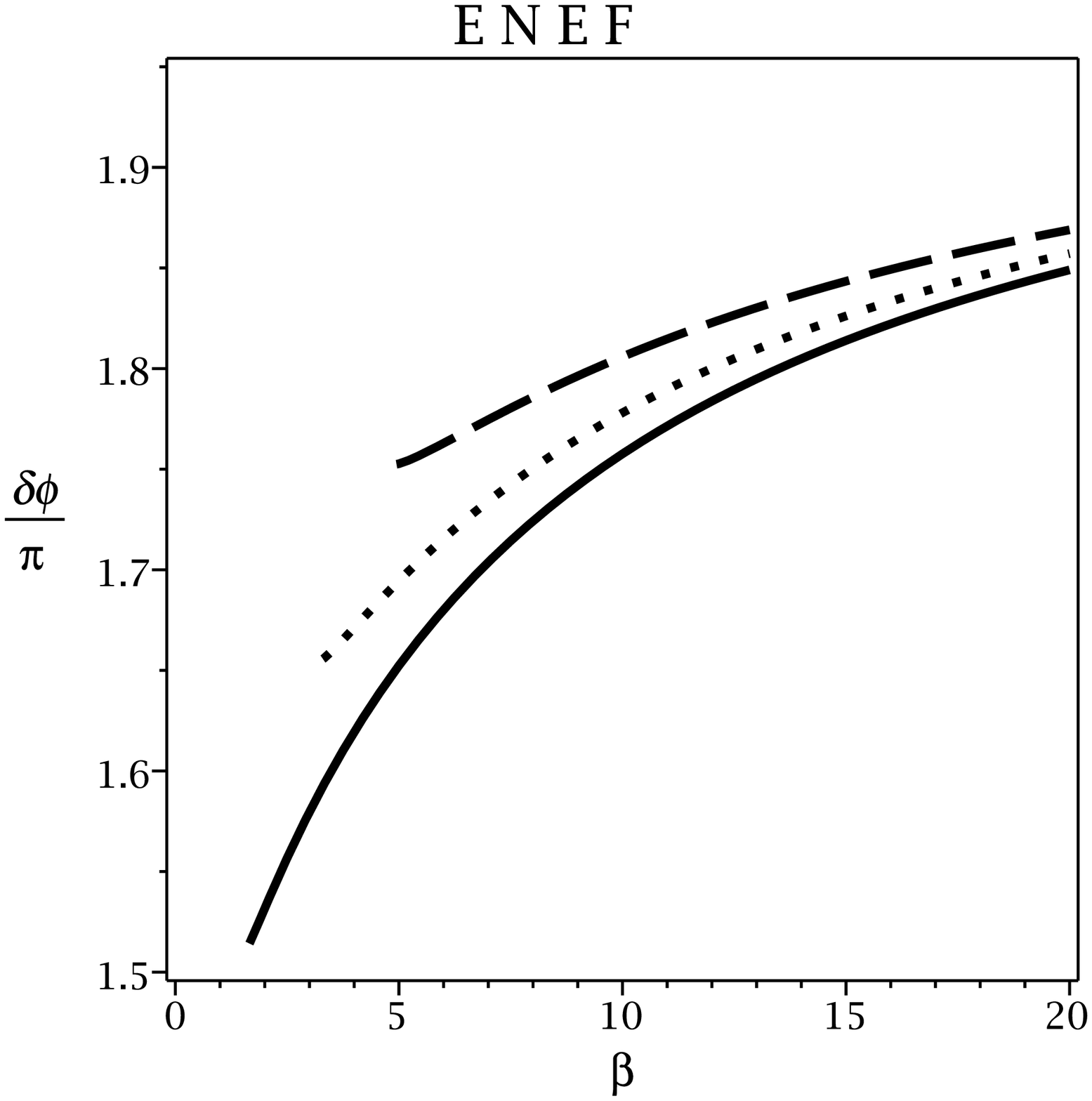} & \epsfxsize=7cm %
\epsffile{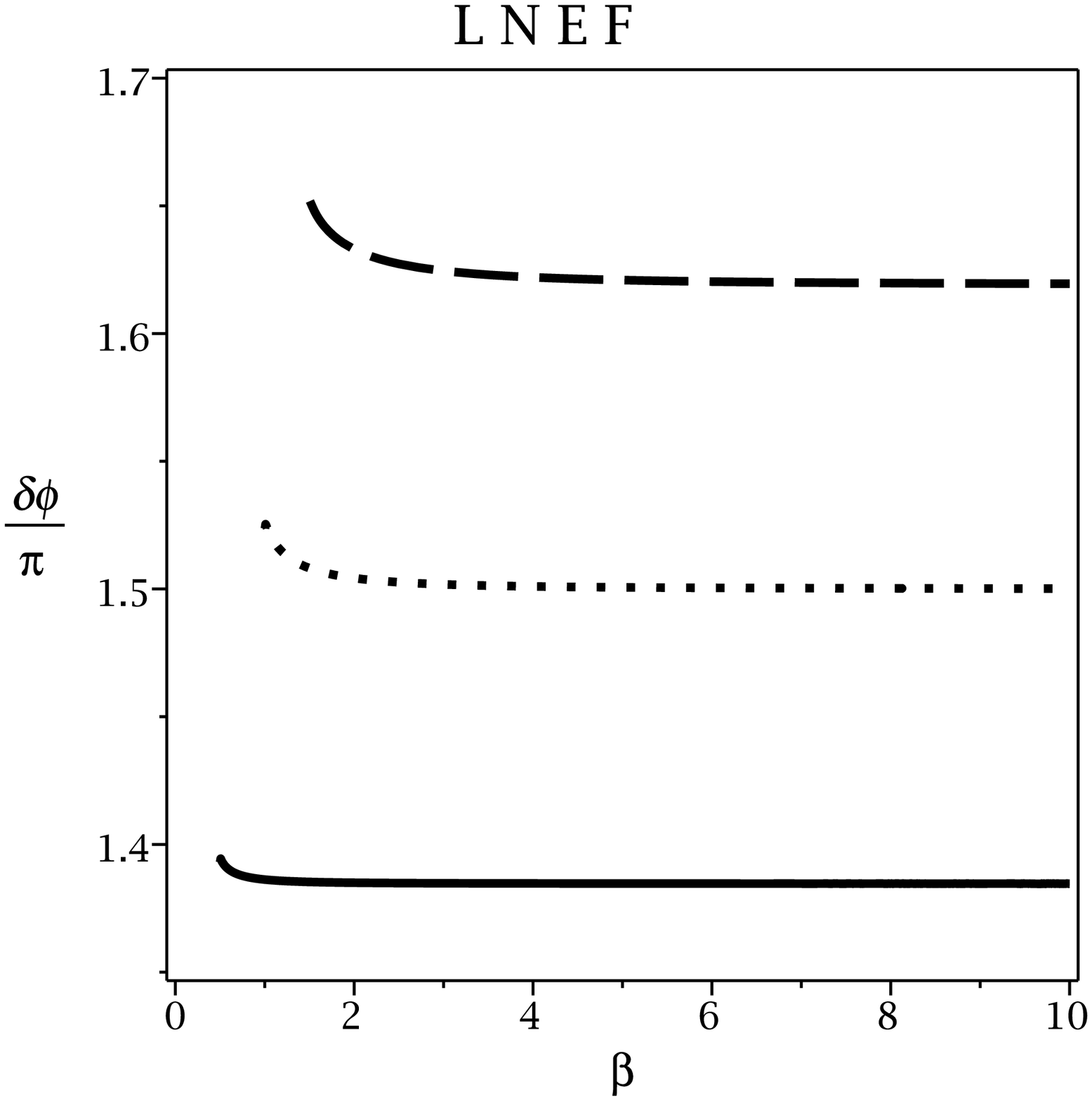} &
\end{array}
$%
\caption{$\delta \phi/\pi$ versus $\protect\beta$ for $d=4$, $l=1$ and $r_{+}=2$%
. \newline Left panel (ENEF): $q=1$ (continuous line), $q=2$
(doted line) and $q=3$ (dashed line). \newline Right panel (LNEF):
$q=1$ (continuous line), $q=2$ (doted line) and $q=3$ (dashed
line).} \label{Fig4GB}
\end{figure}

\begin{figure}[tbp]
$%
\begin{array}{ccc}
\epsfxsize=7cm \epsffile{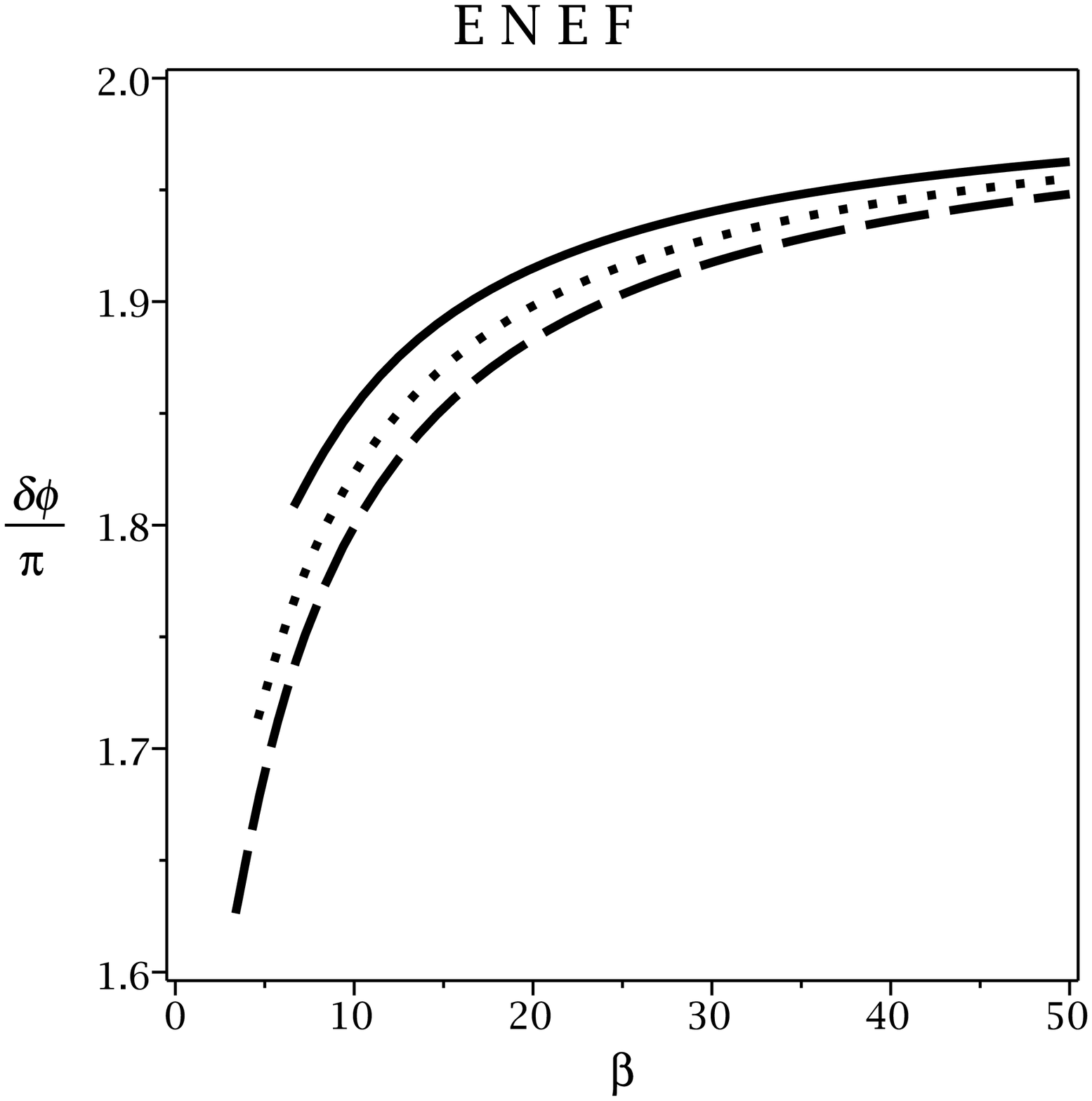} & \epsfxsize%
=7cm \epsffile{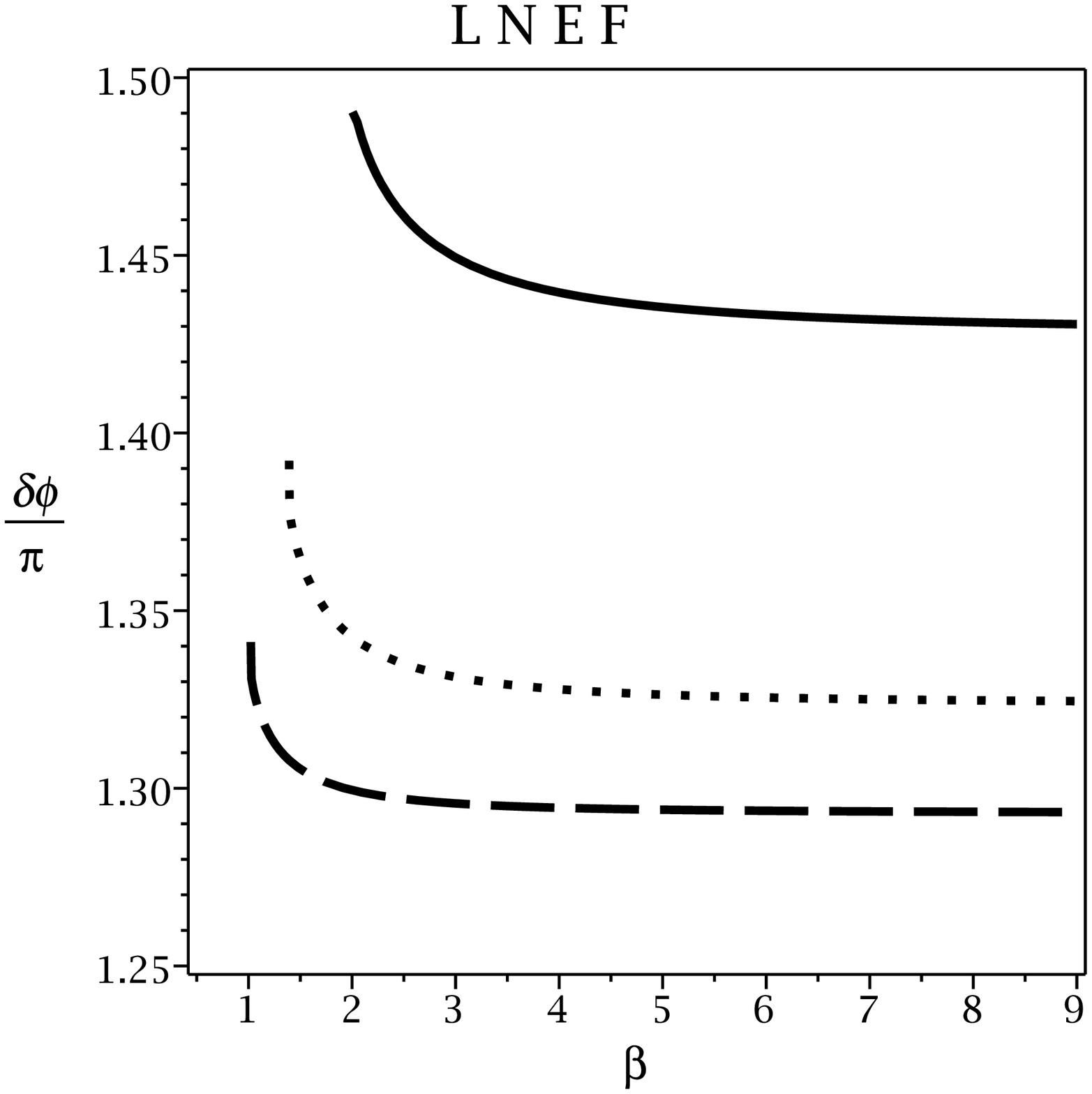} &
\end{array}
$%
\caption{$\delta \phi/\pi$ versus $\protect\beta$ for $d=4$, $l=1$
and $q=1$.
\newline
Left panel (ENEF): $r_{+}=1$ (continuous line), $r_{+}=1.2$ (doted line) and
$r_{+}=1.4$ (dashed line). \newline
Right panel (LNEF): $r_{+}=1$ (continuous line), $r_{+}=1.2$ (doted line)
and $r_{+}=1.4$ (dashed line).}
\label{Fig5GB}
\end{figure}

\begin{figure}[tbp]
$%
\begin{array}{ccc}
\epsfxsize=7cm \epsffile{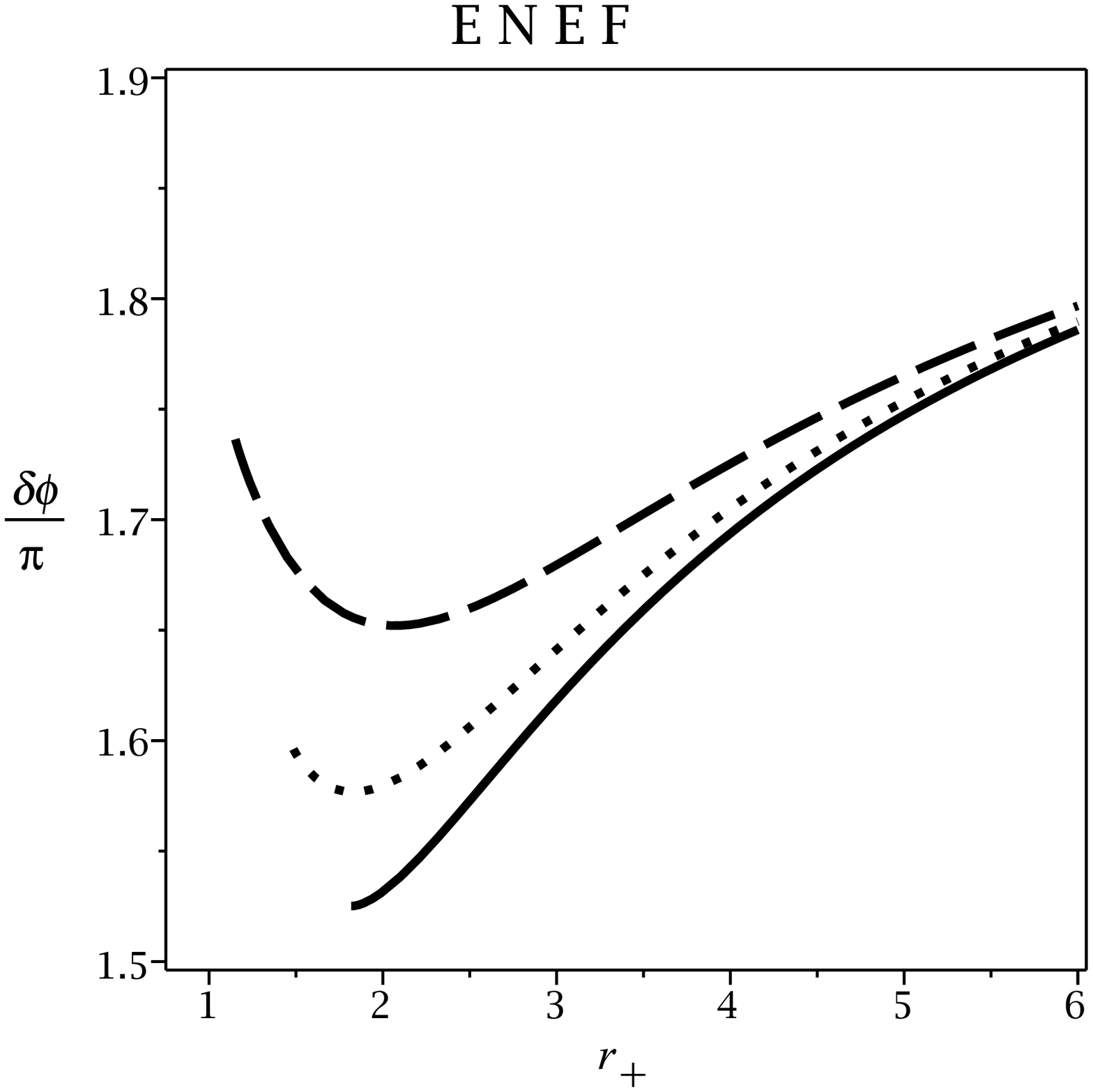} & \epsfxsize%
=7cm \epsffile{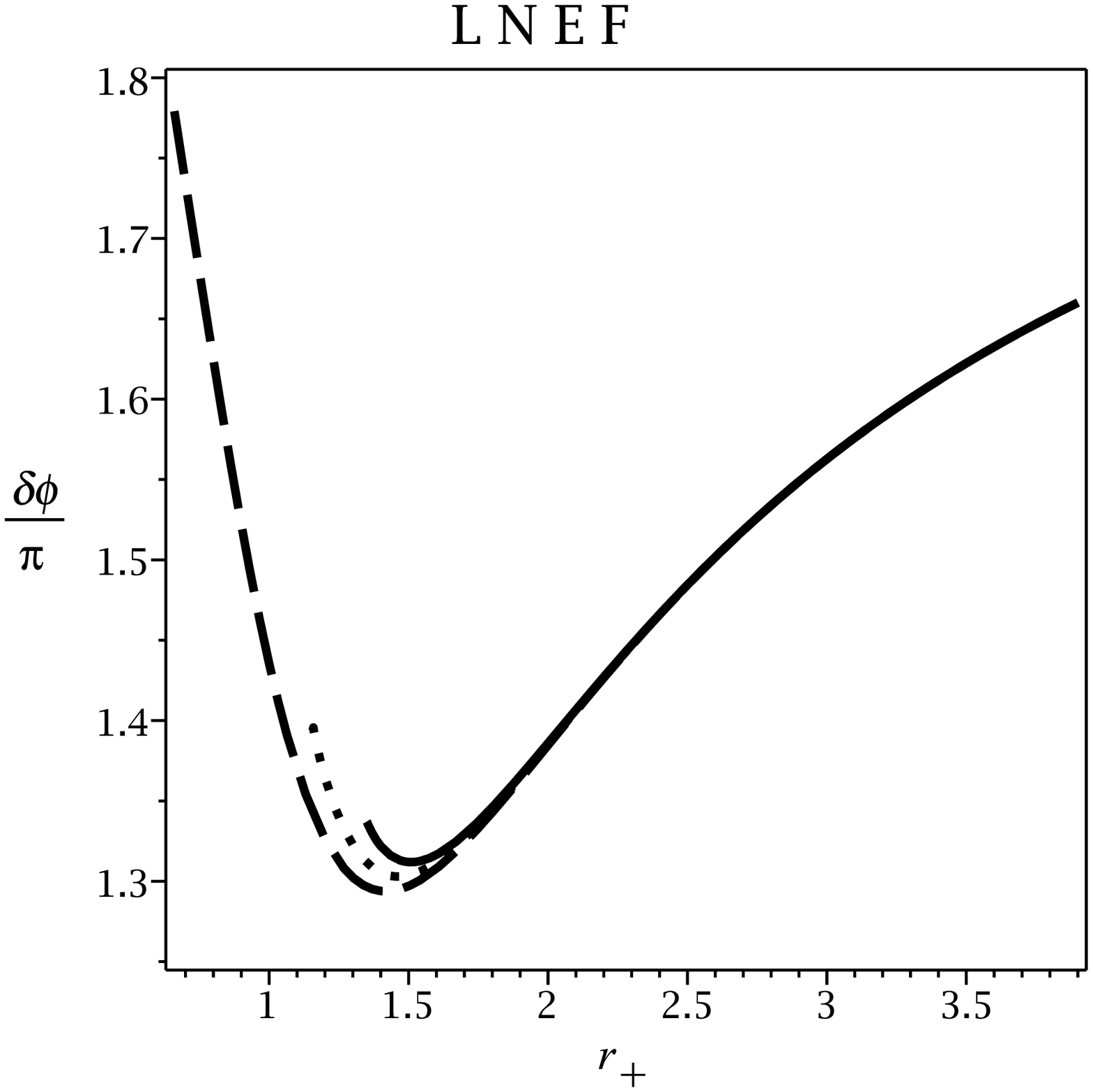} &
\end{array}
$%
\caption{$\delta \phi/\pi$ versus $r_{+}$ for $d=4$%
, $l=2$ and $q=1$. \newline Left panel (ENEF): $\protect\beta=2$
(continuous line), $\protect\beta=3$ (doted line) and
$\protect\beta=5$ (dashed line). \newline
Right panel (LNEF): $\protect\beta=1.1$ (continuous line), $\protect\beta%
=1.5 $ (doted line) and $\protect\beta=5$ (dashed line).}
\label{Fig6GB}
\end{figure}

\begin{figure}[tbp]
$%
\begin{array}{ccc}
\epsfxsize=7cm \epsffile{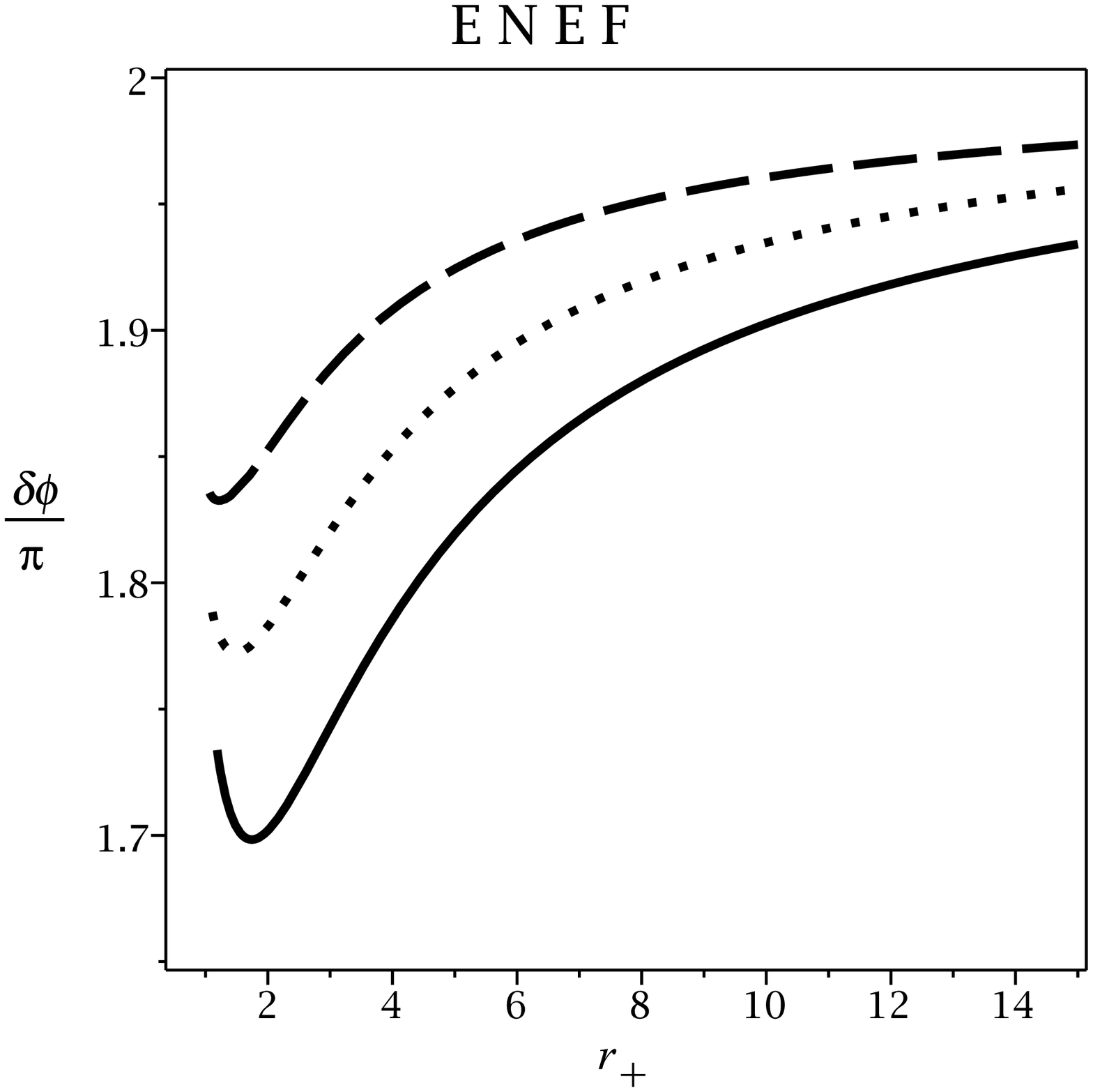} & \epsfxsize%
=7cm \epsffile{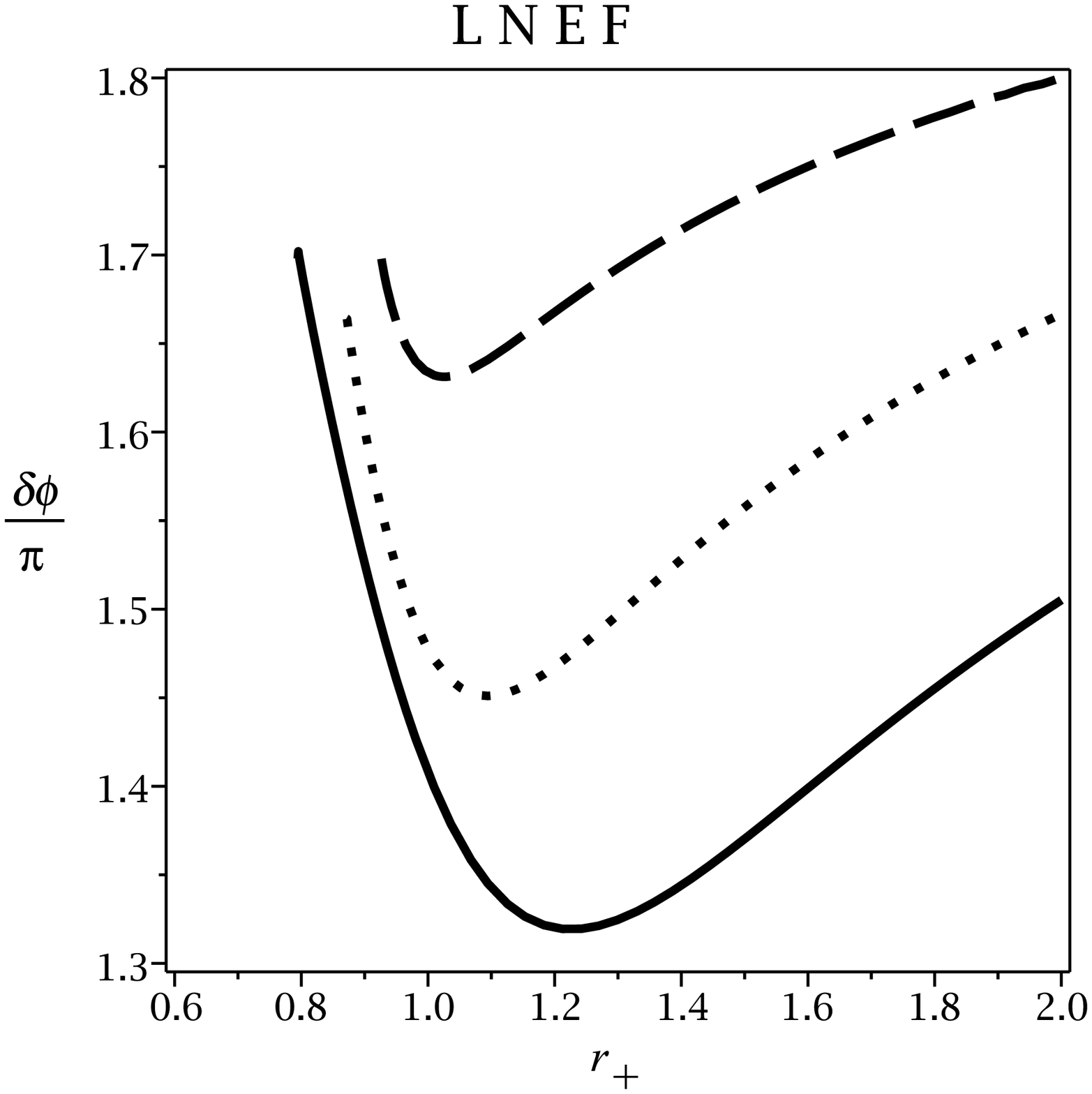} &
\end{array}
$%
\caption{$\delta \phi/\pi$ versus $r_{+}$ for $\protect\beta=4$, $l=1$ and $q=1$%
. \newline Left panel (ENEF): $d=5$ (continuous line), $d=7$
(doted line) and $d=11$ (dashed line). \newline Right panel
(LNEF): $d=5$ (continuous line), $d=7$ (doted line) and $d=11$
(dashed line).} \label{Fig7GB}
\end{figure}

\begin{figure}[tbp]
$%
\begin{array}{ccc}
\epsfxsize=7cm \epsffile{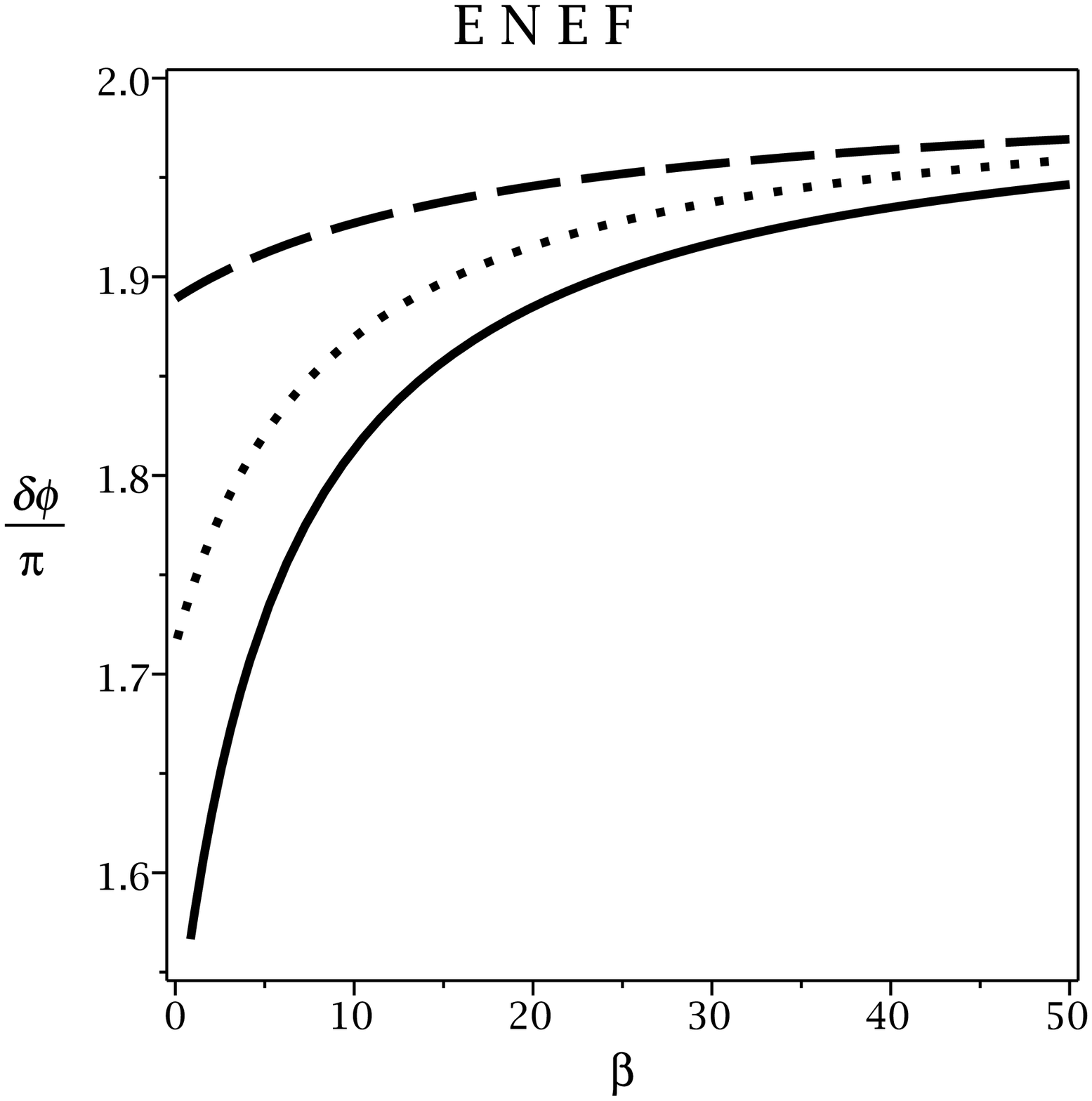} & \epsfxsize%
=7cm \epsffile{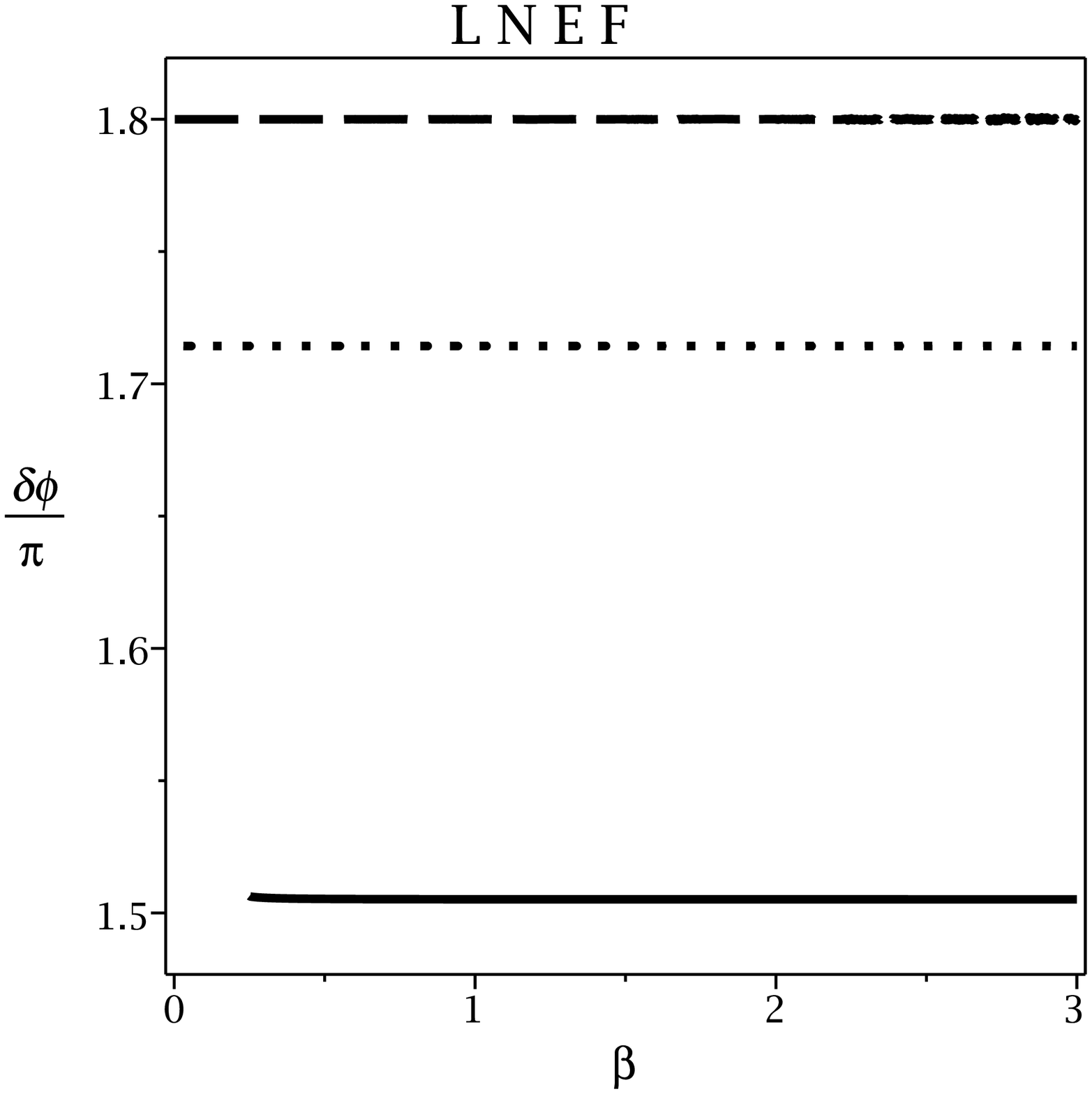} &
\end{array}
$%
\caption{$\delta \phi/\pi$ versus $\protect\beta$ for $r_{+}=2$, $l=1$ and $q=1$%
. \newline Left panel (ENEF): $d=5$ (continuous line), $d=7$
(doted line) and $d=11$ (dashed line). \newline Right panel
(LNEF): $d=5$ (continuous line), $d=7$ (doted line) and $d=11$
(dashed line).} \label{Fig8GB}
\end{figure}

It is obvious that the nonlinearity of electrodynamics can change the value
of deficit angle $\delta \phi $. In order to investigate the effects of
nonlinearity, $r_{+},$ $q$ and dimensionality, we plot $\delta \phi $ versus
$\beta $ and $r_{+}$ (Figs. \ref{Fig4GB}-\ref{Fig8GB}). Figs. \ref{Fig4GB}
and \ref{Fig5GB} show that, for ENEF branch, deficit angle is an increasing
function of nonlinearity parameter while for LNEF branch it is a decreasing
function of $\beta $. In addition, figures of deficit angle show that there
is a minimum for nonlinearity parameter $\beta _{\min }$ in which for $\beta
\leq \beta _{\min }$, the obtained values for deficit angle are not real
(see Figs. \ref{Fig4GB}-\ref{Fig5GB}). Besides, one finds $\beta _{\min }$
increases as the charge parameter of magnetic branes increases, whereas for
increasing value of $r_{+}$, $\beta _{\min }$ decreases (see Figs. \ref%
{Fig4GB} and \ref{Fig5GB}).

The figures of the deficit angle versus $r_{+}$ (see Figs. \ref{Fig6GB} and %
\ref{Fig7GB}) show that there is also a minimum $r_{+_{\min }}$ in which for
$r_{+}\geq r_{+_{\min }}$ the deficit angle is real. For large values of $%
\beta $, the deficit angle is an increasing function of $r_{+}$ (see Figs. %
\ref{Fig6GB} and \ref{Fig7GB}). These figures show that there is an extremum
$r_{+_{ext}}$ that for $r_{+_{min}}\leq r_{+}\leq r_{+_{ext}}$, deficit
angle is a decreasing function of $r_{+}$ whereas for $r_{+}\geq r_{+_{ext}}$
the deficit angle is an increasing function of $\beta$ (see Figs. \ref%
{Fig6GB} and \ref{Fig7GB}).

Considering the fact that obtained results are magnetic branes in arbitrary
dimensions, studying the effect of dimensionality on deficit angle is
another important issue. Figs \ref{Fig7GB} and \ref{Fig8GB} show that for
fixed values of metric parameters, the deficit angle is an increasing
function of $d$. Also, as one can see, $\beta _{min}$ is a decreasing
function of dimensionality and for higher dimensions $\beta _{min}$ goes to
zero (see Fig. \ref{Fig8GB}). Also, numerical analysis confirm that $%
r_{+_{\min }}$ is an increasing function of dimensionality (see Fig. \ref%
{Fig7GB}).

%%%%%%%%%%%%%%%%%%%%%%%%%%%%%%%%%%%%%%%%%%%%%%%%%%%%%%%%%%%%%%%

%%%%%%%%%%%%%%%%%%%%%%%%%%%%%%%%%%%%%%%%%%%%%%%%%%%%%%%%%%%%%%%

%%%%%%%%%%%%%%%%%%%%%%%%%%%%%%%%%%%%%%%%%%%%%%%%%%%%%%%%%%%%%%%
%%%%%%%%%%%%%%%%%%%%%%%%%%%%%%%%%%%%%%%%%%%%%%%%%%%%%%%%%%%%%%%

%%%%%%%%%%%%%%%%%%%%%%%%%%%%%%%%%%%%%%%%%%%%%%%%%%%%%%%%%%%%%%%

%%%%%%%%%%%%%%%%%%%%%%%%%%%%%%%%%%%%%%%%%%%%%%%%%%%%%%%%%%%%%%%

%%%%%%%%%%%%%%%%%%%%%%%%%%%%%%%%%%%%%%%%%%%%%%%%%%%%%%%%%%%%%%%

%%%%%%%%%%%%%%%%%%%%%%%%%%%%%%%%%%%%%%%%%%%%%%%%%%%%%%%%%%%%%%%

%%%%%%%%%%%%%%%%%%%%%%%%%%%%%%%%%%%%%%%%%%%%%%%%%%%%%%%%%%%%%%%
%%%%%%%%%%%%%%%%%%%%%%%%%%%%%%%%%%%%%%%%%%%%%%%%%%%%%%%%%%%%%%%

\section{A class of spinning solutions}

In this section, we generalize the static spacetime to the case of rotating
solutions. As we know, the rotation group in $d$-dimensions is $SO(d-1)$
with $[(d-1)/2]$ independent rotation parameters, in which $[x]$ denotes the
integer part of $x$. The rotating magnetic solutions with $k\leq \lbrack
(d-1)/2]$ rotation parameters may be written as%
\begin{eqnarray}
ds^{2} &=&-\frac{r^{2}+r_{+}^{2}}{l^{2}}\left( \Xi dt-{{\sum_{i=1}^{k}}}%
a_{i}d\phi ^{i}\right) ^{2}+f(r)\left( \sqrt{\Xi ^{2}-1}dt-\frac{\Xi }{\sqrt{%
\Xi ^{2}-1}}{{\sum_{i=1}^{k}}}a_{i}d\phi ^{i}\right) ^{2}  \notag \\
&&+\frac{r^{2}dr^{2}}{(r^{2}+r_{+}^{2})f(r)}+\frac{r^{2}+r_{+}^{2}}{%
l^{2}(\Xi ^{2}-1)}{\sum_{i<j}^{k}}(a_{i}d\phi _{j}-a_{j}d\phi _{i})^{2}+%
\frac{r^{2}+r_{+}^{2}}{l^{2}}dX^{2},  \label{Metric3}
\end{eqnarray}%
where $\Xi =\sqrt{1+\sum_{i}^{k}a_{i}^{2}/l^{2}}$, $dX^{2}$ is the Euclidean
metric on the $(d-k-2)$-dimensional submanifold with volume $V_{d-k-2}$ and $%
f(r)$ is the same as $f(r)$ given in Eqs. (\ref{F11})-(\ref{F33}) for
various gravity. We should note that the non-vanishing components of
electromagnetic field are%
\begin{equation}
F_{rt}=-\frac{(\Xi ^{2}-1)}{\Xi a_{i}}F_{r\phi _{i}}=-\frac{(\Xi ^{2}-1)}{%
\Xi a_{i}}\times \left\{
\begin{array}{cc}
\frac{2ql^{2}}{\left( r^{2}+r_{+}^{2}\right) ^{d_{2}/2}}\exp \left( -\frac{%
L_{W}}{2}\right) , & \text{ENEF}\vspace{0.1cm} \\
\frac{\beta ^{2}\left( r^{2}+r_{+}^{2}\right) ^{d_{2}/2}}{q}\left( 1-\Gamma
\right) , & \text{LNEF}%
\end{array}%
\right. .
\end{equation}

Again, we should note that although this rotating spacetime has no curvature
singularity and horizon, it has a conical singularity at $r=0$.

\subsection{Conserved Quantities \label{Conserve}}

Here, we calculate the angular momentum and mass density of the magnetic
solutions. In order to obtain finite conserved quantities for the
asymptotically AdS solutions, one may use the counterterm method \cite{Mal}.
Here, for the asymptotically AdS solutions of the Lovelock gravity with flat
boundary, $\widehat{R}_{abcd}(\gamma )=0$ (our solutions), the finite energy
momentum tensor is \cite{BY}
\begin{equation}
T^{ab}=\frac{1}{8\pi }\left[ (K^{ab}-K\gamma ^{ab})+2\alpha
_{2}(3J^{ab}-J\gamma ^{ab})+3\alpha _{3}\left( 5P^{ab}-P\gamma ^{ab}\right) +%
\frac{d_{2}}{l_{eff}}\gamma ^{ab}\right] ,  \label{Stress}
\end{equation}%
where $l_{eff}$ is a function of $l$ and $\alpha $, and when $\alpha $ goes
to zero (Einstein solutions), $l_{eff}$ \ reduces to $l$. In Eq. (\ref%
{Stress}), $K^{ab}$ is the extrinsic curvature of the boundary, $K$ is its
trace, $\gamma ^{ab}$ is the induced metric of the boundary, and $J$ and $P$
are, respectively, trace of $J^{ab}$ and $P^{ab}$, where%
\begin{equation}
J_{ab}=\frac{1}{3}%
(K_{cd}K^{cd}K_{ab}+2KK_{ac}K_{b}^{c}-2K_{ac}K^{cd}K_{db}-K^{2}K_{ab}),
\label{Jab}
\end{equation}%
and%
\begin{eqnarray}
P_{ab} &=&\frac{1}{5}%
\{[K^{4}-6K^{2}K^{cd}K_{cd}+8KK_{cd}K_{e}^{d}K^{ec}-6K_{cd}K^{de}K_{ef}K^{fc}+3(K_{cd}K^{cd})^{2}]K_{ab}
\notag \\
&&-(4K^{3}-12KK_{ed}K^{ed}+8K_{de}K_{f}^{e}K^{fd})K_{ac}K_{b}^{c}-24KK_{ac}K^{cd}K_{de}K_{b}^{e}
\notag \\
&&+(12K^{2}-12K_{ef}K^{ef})K_{ac}K^{cd}K_{db}+24K_{ac}K^{cd}K_{de}K^{ef}K_{bf}\}.
\label{Pab}
\end{eqnarray}

In order to compute the conserved charges, one can write the boundary metric
in Arnowitt-Deser-Misner form
\begin{equation}
\gamma ^{ab}dx^{a}dx^{b}=-N^{2}dt^{2}+\sigma _{ij}\left( d\varphi
^{i}+V^{i}dt\right) \left( d\varphi ^{j}+V^{j}dt\right) ,  \label{ADM}
\end{equation}%
where the coordinates $\varphi ^{i}$ are the angular variables
parameterizing the hypersurface of constant $r$ around the origin and $N$
and $V^{i}$ are the lapse and shift functions, respectively. The quasilocal
conserved quantities associated with the stress tensors of Eq. (\ref{Stress}%
) are
\begin{equation}
\mathcal{Q}(\mathcal{\xi )}=\int_{\mathcal{B}}d^{d_{2}}\varphi \sqrt{\sigma }%
T_{ab}n^{a}\mathcal{\xi }^{b},  \label{CC}
\end{equation}%
where $\sigma $ is the determinant of the metric $\sigma _{ij}$, and $n^{a}$
is the timelike unit normal vector to the boundary $\mathcal{B}${\ and }$%
\mathcal{\xi }$ is a Killing vector field. The rotating magnetic spacetime (%
\ref{Metric3}) has two conserved quantities which are associated with the
Killing vectors $\xi =\partial /\partial t$ and $\zeta _{i}=\partial
/\partial \phi ^{i}$. The total mass and angular momentum of the magnetic
brane solutions per unit volume $V_{d-k-2}$, given by
\begin{equation}
M=\int_{\mathcal{B}}d^{d_{2}}x\sqrt{\sigma }T_{ab}n^{a}\xi ^{b}=\frac{(2\pi
)^{k}}{4}\left[ d_{1}(\Xi ^{2}-1)+1\right] m,  \label{M}
\end{equation}%
\begin{equation}
J_{i}=\int_{\mathcal{B}}d^{d_{2}}x\sqrt{\sigma }T_{ab}n^{a}\zeta _{i}^{b}=%
\frac{(2\pi )^{k}}{4}\Xi d_{1}ma_{i},  \label{J}
\end{equation}%
where the mass parameter $m$ comes from the fact that $\lim_{r\rightarrow
0}f(r)=0$. Our last step will be devoted to calculate the electric charge of
the magnetic solutions. To do so, we should consider the projections of the
electromagnetic field tensor on a special hypersurface. The electric charge
per unit volume $V_{d-k-2}$ can be found by calculating the flux of the
electromagnetic field at infinity, yielding
\begin{equation}
Q=\frac{(2\pi )^{k}}{2}ql\sqrt{\Xi ^{2}-1},  \label{Q}
\end{equation}%
which shows that the electric charge is proportional to the magnitude of the
rotation parameters and is zero for the static solutions ($\Xi =1$). This is
due to the fact that the electric field, $F_{tr}$, vanishes for the static
solutions. In addition, since the asymptotically behavior of the
electromagnetic field is the same as that of the Maxwell theory, the
nonlinearity does not affect the total electric charge.

\section{NED as a correction}

It is arguable that, instead of considering nonlinear theories of the
Maxwell field, one can use the method in which the nonlinearity is playing
as a correction term. In other words, one is free to consider nonlinearity
as a perturbation to linear theory and construct a new nonlinear theory.
This treatment is justified with the following reasons. First, in order to
find experimental results for a nonlinear electromagnetic fields, one should
consider its weak nonlinearity and not strong. This is due to the fact that
the Maxwell theory has acceptable consequences in most domains and the
perturbed nonlinear theory of electrodynamics may increases the Maxwell
accuracy. On the other hand, in order to avoid the complexity of nonlinear
theories and obtaining interesting solutions, it is logical to consider the
dominant nonlinearity terms and use them in order to study a nonlinear
theory. As for BI types of nonlinear electrodynamics for large values of
nonlinearity parameter they have same structure with a little differences in
some factors. One can show that the first and second leading order terms
are, respectively, the Maxwell Lagrangian and quadratic power of the Maxwell
invariant. Therefore, in this section we consider following Lagrangian as a
source and study the effects of additional corrction to the Maxwell theory
(MC) as nonlinear electromagnetic field on solutions:
\begin{equation}
L(F)=-F+\eta F^{2}+O(\eta ^{2}).  \label{Lcor}
\end{equation}

One may follow the procedure of previous sections with the mentioned
Lagrangian (\ref{Lcor}) and the metric (\ref{Metric3}) to obtain%
\begin{equation}
F_{rt}=-\frac{(\Xi ^{2}-1)}{\Xi a_{i}}F_{r\phi _{i}}=-\frac{(\Xi ^{2}-1)}{%
\Xi a_{i}}\left( \frac{q}{\left( r^{2}+r_{+}^{2}\right) ^{\frac{d_{4}}{2}}}-%
\frac{4q^{3}\eta }{l^{2}\left( r^{2}+r_{+}^{2}\right) ^{\frac{3d_{2}}{2}}}%
\right) .  \label{Ftrcor}
\end{equation}

Inserting Eq. (\ref{Ftrcor}) in the gravitational field equations, we find
the following metric functions for the EN, the GB and the TOL gravities in
the presence of Lagrangian (\ref{Lcor})
\begin{equation}
f_{EN}=\frac{2Ml^{3}}{\left( r^{2}+r_{+}^{2}\right) ^{d_{3}/2}}-\frac{%
2\left( r^{2}+r_{+}^{2}\right) }{d_{1}d_{2}}\left( \Lambda -\frac{%
4d_{1}l^{2}q^{2}}{d_{3}\left( r^{2}+r_{+}^{2}\right) ^{d_{2}}}+\frac{%
32d_{1}l^{4}q^{4}\eta }{\left( 3d-7\right) \left( r^{2}+r_{+}^{2}\right)
^{2d_{2}}}\right) ,  \label{F111}
\end{equation}%
\begin{equation}
f_{GB}=\frac{\left( r^{2}+r_{+}^{2}\right) }{2d_{3}d_{4}\alpha _{2}}\left(
1-\Psi ^{\frac{1}{2}}\right) ,  \label{F222}
\end{equation}%
\begin{equation}
f_{TOL}=\frac{\left( r^{2}+r_{+}^{2}\right) }{d_{3}d_{4}\alpha _{2}}\left(
1-\Psi ^{\frac{1}{3}}\right) ,  \label{F333}
\end{equation}%
where%
\begin{equation}
\Psi =1+\frac{2\chi d_{3}d_{4}\alpha _{2}}{d_{1}d_{2}}\left( \Lambda -\frac{%
d_{1}d_{2}l^{3}M}{\left( r^{2}+r_{+}^{2}\right) ^{d_{1}/2}}-\frac{%
4d_{1}l^{2}q^{2}}{d_{3}\left( r^{2}+r_{+}^{2}\right) ^{d_{2}}}+\frac{%
32d_{1}l^{4}q^{4}\eta }{(3d_{2}-1)\left( r^{2}+r_{+}^{2}\right) ^{2d_{2}}}%
\right) ,
\end{equation}%
and $\chi =4$ and $3$ for the GB theory and the TOL gravity, respectively.

We should note that, regardless of various coefficients, one can obtain
these solutions, directly, by suitable series expansions of Eqs. (\ref{F11})
-- (\ref{F33}). In addition, in agreement with Eqs. (\ref{M}), (\ref{J}) and
(\ref{Q}), independent calculations show that the conserved charges do not
depend on the nonlinearity parameter of BI type NED theories.

Here, we are in position to study the deficit angle. To do so, we employ the
method that was mentioned in previous sections and plot various appropriate
graphs. It is a matter of calculation to show that the second order
derivation of the metric with respect to radial coordinate will be in the
following form for all mentioned gravity branched

\begin{equation}
\left. \frac{d^{2}f(r)}{dr^{2}}\right\vert _{r=0}=-\frac{2\Lambda }{d_{2}}-%
\frac{8l^{2}q^{2}}{d_{2}r_{+}^{2d_{2}}}+\frac{64l^{4}q^{4}}{%
d_{2}r_{+}^{4d_{2}}}\eta +O(\eta ^{2}),  \label{dff}
\end{equation}%
where confirms that deficit angle does not depend on the Lovelock
coefficient.

%%%%%%%%%%%%%%%%%%%%%%%%%%%%%%%%%%%%%%%%%%%%%%%%%%%%%%%%%%%%%%%%%%%%%%%%%%%%%%%%%%%%%%%%%%%%%%%%%%%%%%%%%
\begin{figure}[tbp]
$%
\begin{array}{ccc}
\epsfxsize=7cm \epsffile{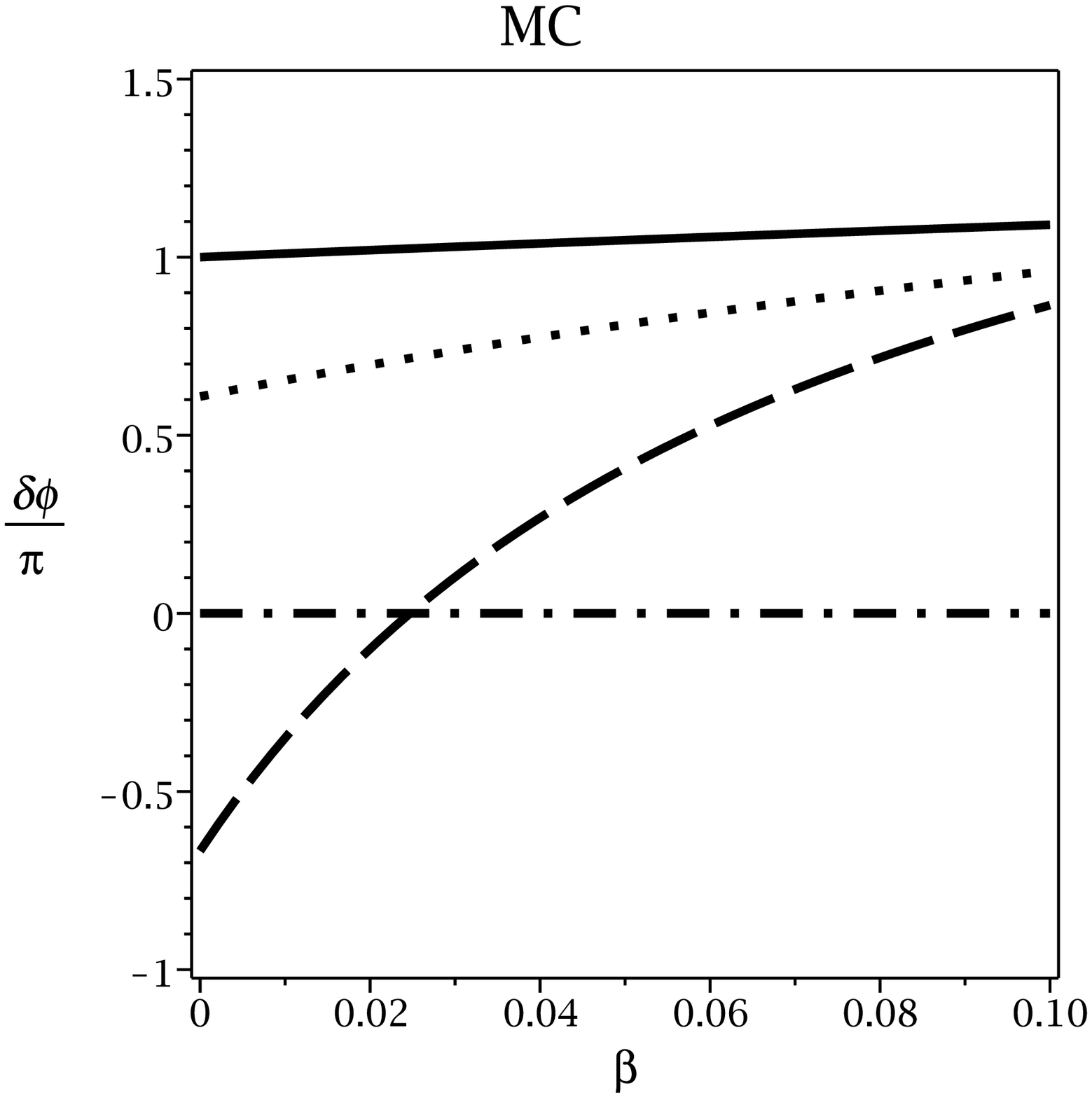} & %
\epsfxsize=7cm \epsffile{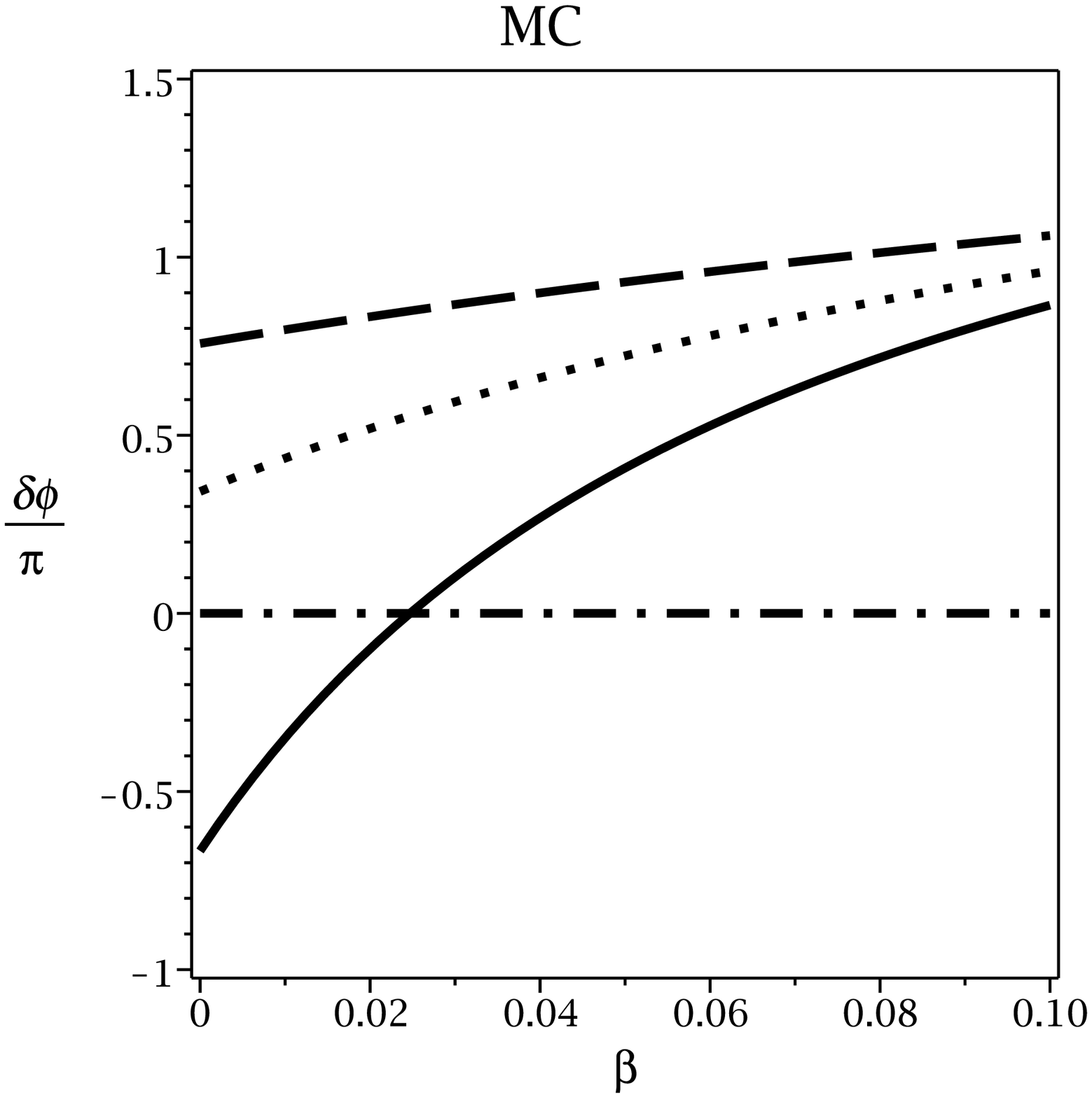} &
\end{array}
$%
\caption{$\delta \phi/\pi$ versus $\protect\eta$ for $l=1$ and
$d=4$. \newline
Left panel: $r_{+}=2$, $q=2$ (continuous line), $q=2.5$ (doted line) and $%
q=3 $ (dashed line). \newline
Right panel: $q=3$, $r_{+}=2$ (continuous line), $r_{+}=2.1$ (doted line)
and $r_{+}=2.2$ (dashed line).}
\label{Fig10GB}
\end{figure}
%%%%%%%%%%%%%%%%%%%%%%%%%%%%%%%%%%%%%%%%%%%%%%%%%%%%%%%%%%%%%%%

%%%%%%%%%%%%%%%%%%%%%%%%%%%%%%%%%%%%%%%%%%%%%%%%%%%%%%%%%%%%%%%
\begin{figure}[tbp]
$%
\begin{array}{ccc}
\epsfxsize=7cm \epsffile{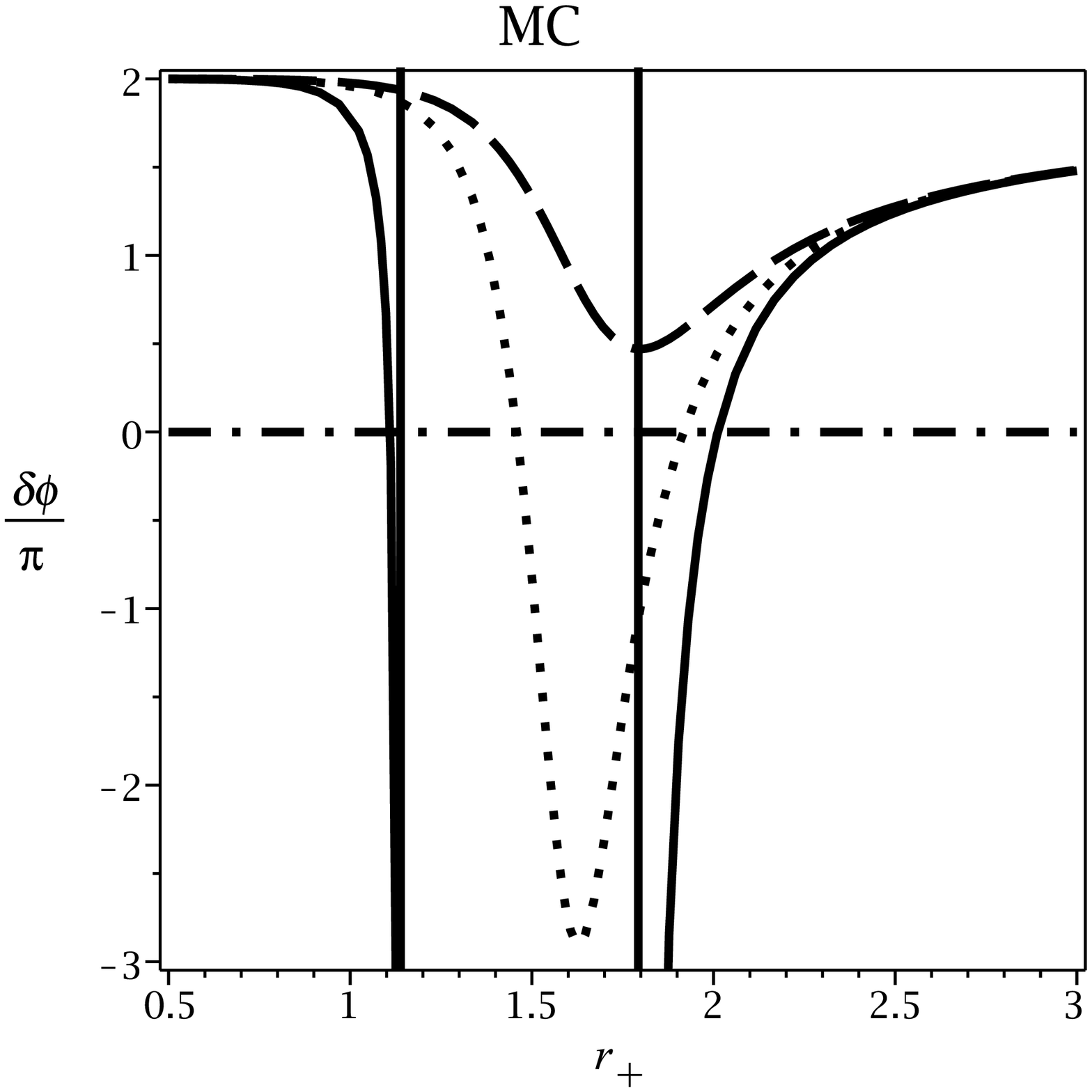} & %
\epsfxsize=7.2cm
\epsffile{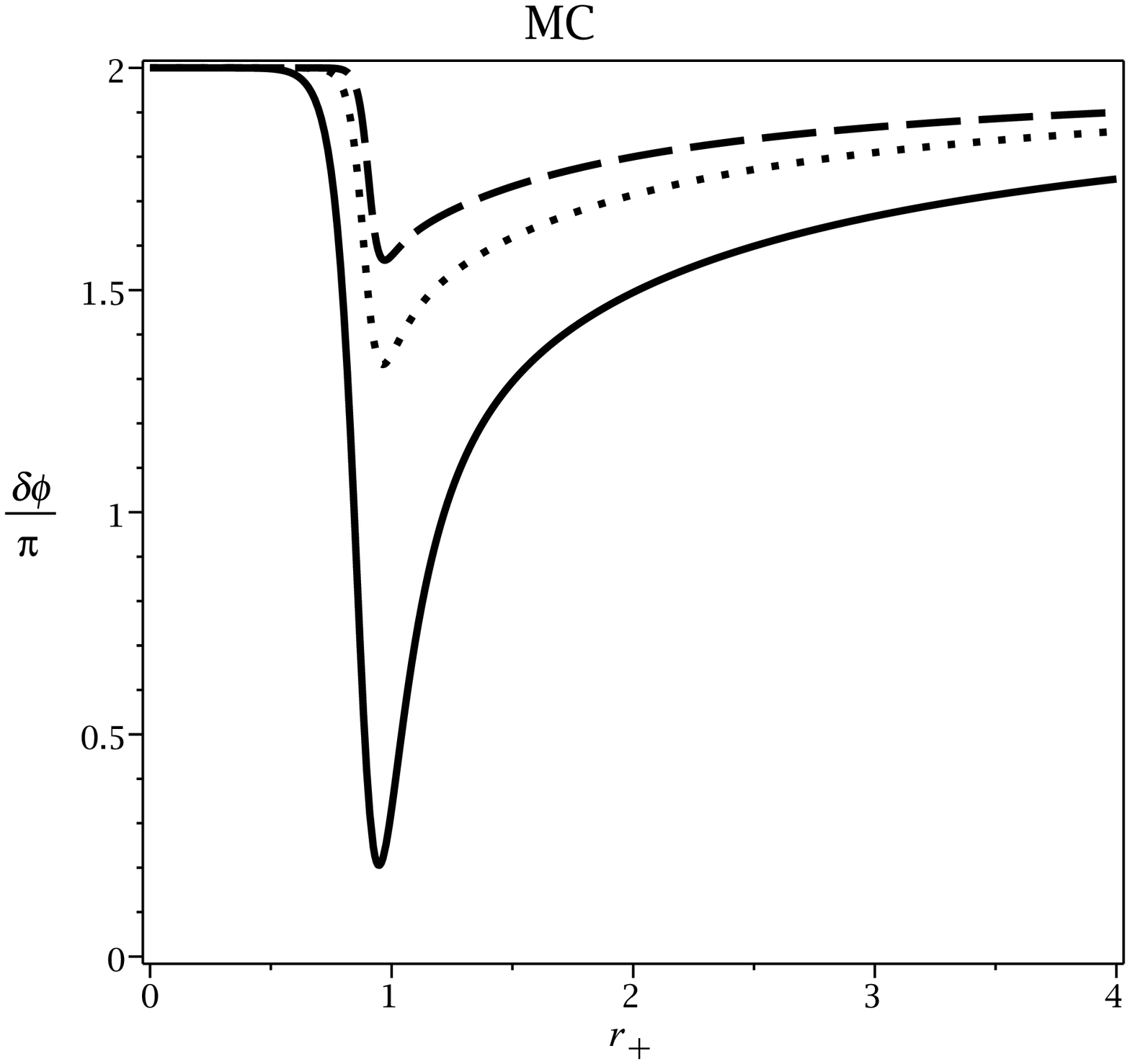} &
\end{array}
$%
\caption{$\delta \phi/\pi$ versus $r_{+}$ for $l=1$.
\newline Left panel: $q=3$, $d=4$, $\protect\eta=0.02$ (continuous line),
$\protect\eta=0.05$ (doted line) and $\protect\eta=0.08$ (dashed
line). \newline Right panel: $q=1$, $\protect\eta=0.05$, $d=5$
(continuous line), $d=8$ (doted line), $d=11$ (dashed line).}
\label{Fig11GB}
\end{figure}
%%%%%%%%%%%%%%%%%%%%%%%%%%%%%%%%%%%%%%%%%%%%%%%%%%%%%%%%%%%%%%%

Studying the effects of charge parameter show that, for very small values of
$q$ and $\eta =0$, calculated deficit angle is nonzero and is a decreasing
function of charge (see left panel in Fig. \ref{Fig10GB} ). As charge
increases, for certain range of correction parameter, deficit angle is
negative and there is a $\eta _{0}$ where calculated deficit angle is zero.
This $\eta _{0}$ is an increasing function of charge (see left panel in Fig. %
\ref{Fig10GB} ). As for the effects of $r_{+}$, plotted graphs have similar
behavior as charge whereas the effects of $r_{+}$ are exactly opposite of
the effects of charge (see right panel in Fig. \ref{Fig10GB} ).

Considering different values of correction parameter, the deficit angle
versus $r_{+}$ shows that, calculated deficit angles have different
behaviors. For small values of nonlinearity, three different behaviors are
seen for different regions of $r_{+}$ in which these regions are specified
with $r_{+_{Div1}}$ and $r_{+_{Div2}}$ (see left panel in Fig. \ref{Fig11GB}
). For $0<r_{+}<r_{+_{Div1}}$, deficit angle is an decreasing function of $%
r_{+}$ and in $r_{+}=r_{+_{Div1}}$, there is a divergency. In this region
calculated deficit angles are positive and real valued and in case of $%
r_{+_{Div1}}<r_{+}<r_{+_{Div2}}$ for calculated deficit angle, first it is a
decreasing and then increasing function of $r_{+}$ and for $%
r_{+}=r_{+_{Div2}}$ second divergency happens. Next, for $r_{+}>r_{+_{Div2}}$%
, one finds that deficit angle is an increasing function of $r_{+}$ but
there exists a region in which calculated deficit angles are negative and
for an $r_{+_{0}}$ deficit angle is zero (see left panel in Fig. \ref%
{Fig11GB} ).

\begin{figure}[tbp]
$%
\begin{array}{ccc}
\epsfxsize=7cm \epsffile{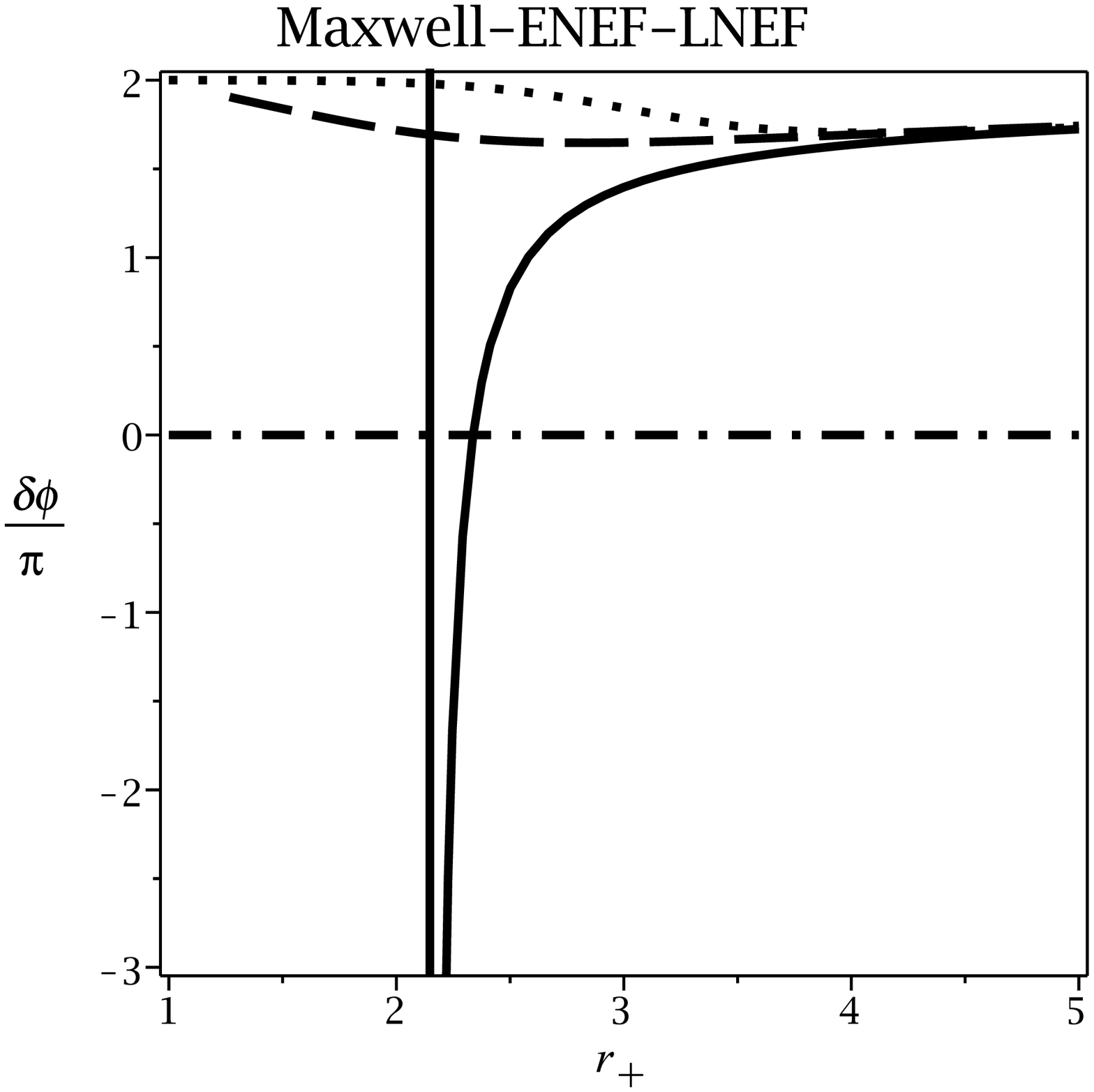} & \epsfxsize=7cm %
\epsffile{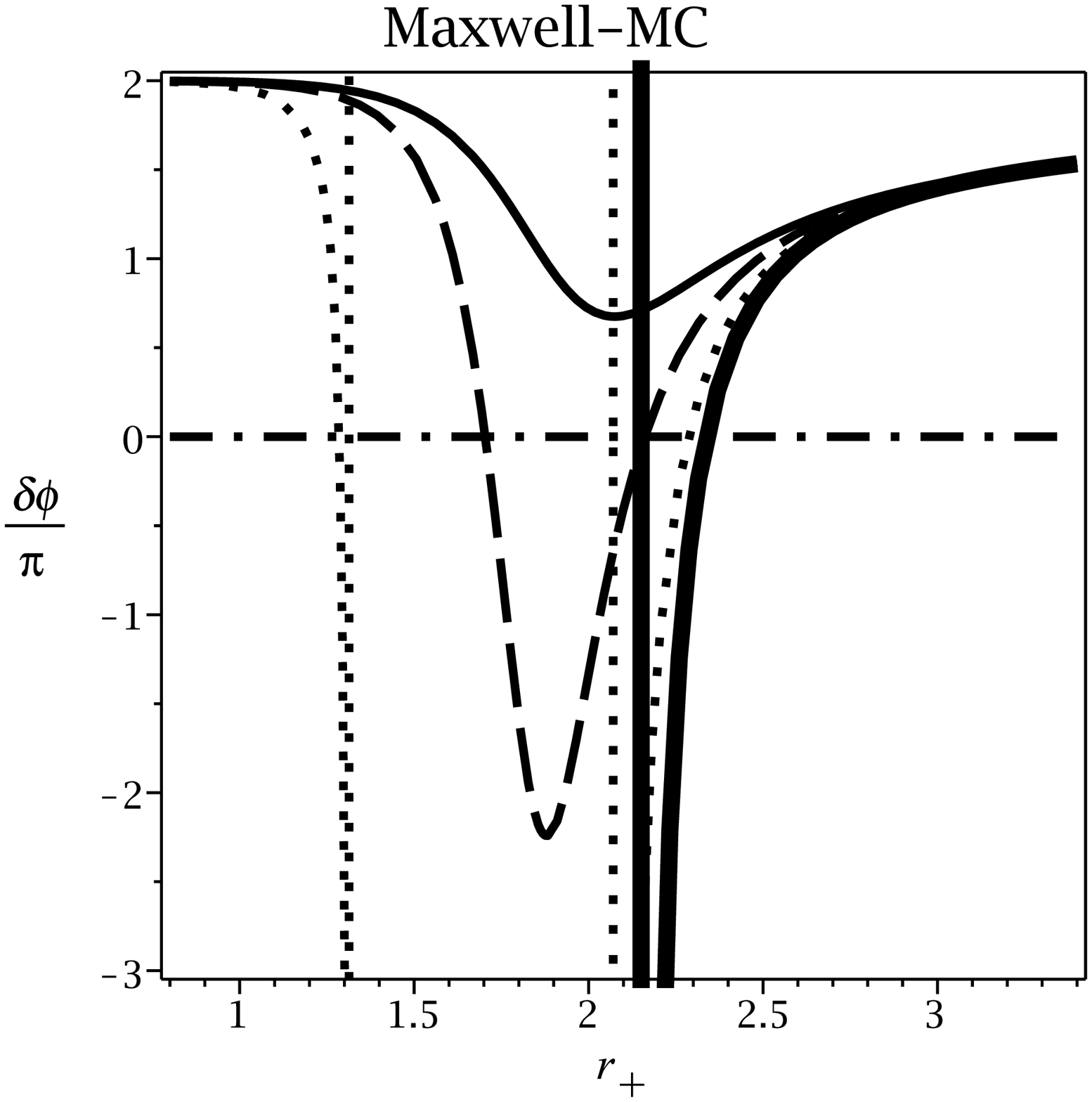} &
\end{array}
$%
\caption{$\delta \phi/\pi$ versus $r_{+}$ for $l=1$, $q=4$ and
$d=4$. \newline Left panel: Maxwell case (continuous line), ENEF
case for $\protect\beta=5$ (doted line) and LNEF case for
$\protect\beta=5$ (dashed line), respectively.
\newline
Right panel: Maxwell case (bold line), MC case for $\protect\eta=0.02 $
(doted line), Maxwell Correction case for $\protect\eta=0.05$ (dashed line)
and MC case for $\protect\eta=0.08$ (continuous line).}
\label{Fig13GB}
\end{figure}

Next, for larger values of nonlinearity, there are regions identified with
specific values naming $r_{+_{1}}$, $r_{+_{ext}}$ and $r_{+_{2}}$. For $%
0<r_{+}<r_{+_{1}}$, deficit angles for different values of nonlinearity
parameter are almost the same. In other words, calculated values of deficit
angle are almost independent of variation of nonlinearity parameter because
its effect is so small. $r_{ext}$ is an extremum in which for $r_{+_{1}}\leq
r_{+}\leq r_{+_{ext}}$, deficit angle decreases where $r_{+}$ increases
while for $r_{+_{2}}\geq r_{+}\geq r_{+_{ext}}$, deficit angle is an
increasing function of $r_{+}$ (see Fig. \ref{Fig13GB} ). Finally for large
values of $r_{+}$ ($r_{+_{2}}\leq r_{+}$) similar behavior as for case of
small values of $r_{+}$ ($0<r_{+}<r_{+_{1}}$) is observed. Calculated values
of deficit angle are almost independent of nonlinearity parameter and are
almost the same. $r_{+_{1}}$, $r_{+_{2}}$, $r_{+_{ext}}$ and related deficit
angle to this extremum are increasing functions of nonlinearity parameter.
As for the effects of dimensions, it is evident from plotted graphs that $%
r_{+_{ext}}$ (and related deficit angle) is a decreasing (increasing)
function of dimensions (see right panel in Fig. \ref{Fig11GB} ). These
figures indicate that there exist regions, in which calculated values of
deficit angle for different dimensions lead to almost same result and it is
almost independent of dimensions.

Here, we present a geometric interpretation for negative deficit
angle. Considering a two dimensional plain, we can cut segment of
a certain angular size and then sew together the edges to obtain a
conical surface. The deleted segment from the plan is known as
deficit angle with positive values. Now, we imagine a new
situation when a segment is added to a new plane to obtain a flat
surface with a saddle-like cone (for more details one can see Fig.
$2$ in Ref. \cite{LR}). This added segment is corresponding to a
negative deficit angle (or surplus angle) \cite{LR,NegDA}. We
should mention that although the deleted segment is bounded by the
value of $2\pi$ the added segment is unbounded. Therefore, we
conclude that the range of deficit angles is from $-\infty$ to
$2\pi$.

%%%%%%%%%%%%%%%%%%%%%%%%%%%%%%%%%%%%%%%%%%%%%%%%%%%%%%%%%%%%%%%
%%%%%%%%%%%%%%%%%%%%%%%%%%%%%%%%%%%%%%%%%%%%%%%%%%%%%%%%%%%%%%%

\section{Closing Remarks}

In this paper, we supposed that the geometry and matter field of spacetime
come from the Lovelock gravity and NED. At first, we considered a suitable
static metric to find horizonless magnetic solutions. We found that for
having the real electromagnetic field, we should consider a lower bound ($%
\rho_{0}$) for the coordinate $\rho $. We discussed the geometric
properties of EN, GB and TOL solutions and found that although
these solutions have no curvature singularity, there is a conical
singularity at $r =0$ with a deficit angle $\delta \phi =8\pi \mu
$, where one can interpret $\mu $ as the mass per unit volume of
the magnetic brane. In addition, we found that both the NED and
the Lovelock gravity do not affect the asymptotically behavior of
the solutions, and in other words, obtained solutions are
asymptotically AdS. We obtained deficit angle of the conical
geometry and investigated the effects of Lovelock gravity and NED.
At first, calculated values for deficit angle showed that it is
independent of the GB and the TOL parameters. In other words, we
found that the Lovelock parameters do not affect the deficit
angle. This result comes from the fact that the value of second
derivatives of metric function does not depend on the Lovelock
coefficients which is the consequence of geometric properties of
$t=constant$ and $r=constant$ hypersurface (this hypersurface is a
Ricci flat manifold). This behavior is similar to property of
Ricci-flat black holes in higher orders of the Lovelock gravity,
in which their horizons and conserved quantities of black hole do
not depend on the Lovelock parameters.

We also investigated the effects of nonlinear electrodynamics.
Although both ENEF and LNEF branches are BI type, they have
different nature. We found that there is a minimum value for the
nonlinearity parameter where for $\beta \leq \beta _{\min }$ the
deficit angle was not real. This is because of the behavior of
Lambert function which is present in ENEF branch and the
logarithmic function which is appeared in LNEF branch. We also
showed that considering higher dimensional solutions, $\beta
_{\min }$ may be changed and for certain dimensions, the deficit
angle is real for arbitrary $\beta $ ($\beta_{\min} <0$). We found
that the deficit angle is an increasing function of nonlinearity
parameter in ENEF whereas for LNEF it showed opposite behavior. We
also saw that increasing the charge parameter leads to increasing
$\beta _{\min }$ while for increasing $r_{+}$, the value of $\beta
_{\min }$ decreased.

Looking at the behavior of deficit angle versus $r_{+}$, we found that there
is an $r_{+_{\min }}$ where for $r_{+} \geq r_{+_{\min }}$, the deficit
angle is real valued. Moreover, we found that for small values of the
nonlinearity parameter, the deficit angle is only an increasing function of $%
r_{+}$ whereas for increasing value of $\beta$ there will be $r_{+_{ext}}$
in which for $r_{+_{\min }} \leq r_{+} \leq r_{+_{ext}}$ the deficit angle
is a decreasing function of $r_{+}$ and for $r_{+} \geq r_{+_{ext}}$ it
increases as $r_{+}$ increases.

Next step was devoted to introduce spinning magnetic branes which is
horizonless. We found that for rotating magnetic branes there is an electric
field in addition to the magnetic one. We employed the Gauss law and the
counterterm method to calculate the electric charge, finite mass and angular
momentum of rotating magnetic brane solutions. We found that the electric
charge is proportional to the rotation parameters and it vanishes for the
static solutions ($\Xi =1$). We should note that vanishing the electric
charge for $\Xi =1$ is due the fact that the electric field, $F_{tr}$,
vanishes for the static solutions.

As one can see for the weak nonlinearity power, the obtained deficit angle
for different theories of nonlinearity has different values comparing to the
Maxwell theory. One may argue that, for large values of $\beta$, the
obtained values for deficit angle, should lead to those of the Maxwell
theory and support this statement with fact that for large values of $\beta$%
, these two electromagnetic fields become Maxwell theory. This
idea is an acceptable one, when we are only dealing with the
electromagnetic fields. But in calculation of deficit angle, we
are using the second derivation of metric function. Due to
different structures of nonlinear theories (logarithmic and
exponential ones), it is most likely that this property of these
two nonlinear electrodynamics (for large values of $\beta $, they
lead to the Maxwell theory) is not preserved and therefore, the
obtained values are different. In other words, one may expect to
see different values for deficit angle even for large values of
nonlinearity parameter and they are not necessarily the same as
Maxwell ones. It means that, although these two types of nonlinear
theory are BI type and for $\beta \longrightarrow \infty$ they
lead to same result, they are completely different theories with
their different characteristics and properties.

In addition we found that plotted graph for the Maxwell theory,
presents a divergency which is due to root(s) of $f^{\prime \prime
}$. While for considering nonlinear theories, the divergency
vanishes and calculated values of deficit angle and plotted graphs
showed no divergency. In other words, in process of going from
linear theory (Maxwell) to a nonlinear theory (logarithmic form or
exponential one), calculated values of deficit angle will be
divergence free and it has smooth behavior. Therefore, considering
nonlinear theories, changes properties of solutions and solve the
problems regarding the linear theory which is of the primary
motivation of considering nonlinear electrodynamics. It is
notable, that considering nonlinear theories put some restriction
on values. In other words, there is a region in which the
calculated values of deficit angle are not real. But this region
is not where the divergency of the Maxwell theory exists. In other
words, by considering suitable value of nonlinearity parameter,
one can cover regions in which the Maxwell theory has divergency.

Later, we investigated the effects of nonlinearity as a
correction. We found that this theory despite other two nonlinear
theories (logarithmic and exponential ones) is always real value
and there is no region in which deficit angle is imaginary.
Plotted graphs of this theory also showed that,
variation of nonlinearity parameter is only effective in a region ($%
r_{+_{1}}\leq r_{+}\leq r_{+_{2}}$) and in other regions
($r_{+_{1}}\geq r_{+}$, $r_{+}\geq r_{+_{2}}$), it is almost
independence of this variation. Same behavior was seen for the
effects of dimensions as well. It was shown that, the construction
of this theory is in a way that for small and large values of
$r_{+}$ the effect of nonlinearity part, decreases rapidly and
almost vanishes and the structure of magnetic branes (cone-like)
is similar to the Maxwell theory and it is as if there is no
correction part. On the other hand for small values of correction
parameter, not only it did not solve divergency of the Maxwell
field, but it also added another divergency to it. In other words,
two divergencies in the case of very weak correction parameter
were seen in MC theory. This shows the fact that, this theory of
nonlinearity and its deficit angle are quite sensitive to
modification of correction parameter. This sensitivity is stronger
even for small values of correction parameter. Although for some
regions the calculated values of deficit angle are almost same as
the one for the Maxwell theory, there is an effective range in
which nonlinearity (correction) will be dominant and has the most
contribution in deficit angle.

Other interesting issues in the results were existence of the
negative, root and divergencies for deficit angle. Positive
deficit angle is representing a cone like structure for the
object. Whereas the negative deficit angle is denoted as extra
angle which is know as surplus angle \cite{LR,NegDA}. This extra
angle changes the shape of the object into a saddle like cone.

Finally it is worthwhile to think about the physical properties
deficit angle as well as surplus one. In addition, One may
investigate the possible wormhole solutions \cite{Worm} of the
mentioned models, This works are under examination.

\begin{acknowledgements}
We would like to thank the anonymous referees for valuable
suggestions. We thank Shiraz University Research Council. This
work has been supported financially by the Research Institute for
Astronomy and Astrophysics of Maragha, Iran.
\end{acknowledgements}

\end{document}